\documentclass[amsmath,amssymb,aps,prl,superscriptaddress,twocolumn]{revtex4-2}
\usepackage{amsmath}
\usepackage{graphicx}
\usepackage[table]{xcolor}

\begin{document}


\title{First-principles calculation of coherence length and penetration depth based on density functional theory for superconductors}


\author{Mitsuaki Kawamura}
\email[]{kawamura-mitsuaki-zr@ynu.ac.jp}
\homepage[]{https://mkawamura.ynu.ac.jp/}
\affiliation{Division of Intelligent Systems Engineering, Yokohama National University, Yokohama 240-8501, Japan}

\author{Takuya Nomoto}
\affiliation{Department of Physics, Tokyo Metropolitan University, Hachioji 192-0397, Japan}

\author{Niklas Witt}
\affiliation{Institute for Theoretical Physics and Astrophysics and W\"{u}rzburg-Dresden Cluster
 of Excellence ct.qmat, University of W\"urzburg, W\"urzburg 97074, Germany}

\author{Ryotaro Arita}
\affiliation{Department of Physics, University of Tokyo, Hongo, Tokyo 113-0033, Japan}
\affiliation{Center for Emergent Matter Science, RIKEN, Wako 351-0198, Japan}

\date{\today}

\begin{abstract}
We develop a first-principles framework for evaluating the fundamental length scales of superconductivity, namely the coherence length $\xi_0$ and the magnetic penetration depth $\lambda_\mathrm{L}$, within superconducting density functional theory (SCDFT). By incorporating finite-momentum Cooper pairs, we formulate a microscopic scheme that enables a consistent and parameter-free determination of $\xi_0$, $\lambda_\mathrm{L}$, and the superconducting transition temperature $T_\mathrm{c}$ on the same theoretical footing. Applying the method to representative elemental superconductors, the A15 compound V$_3$Si, and H$_3$S under high pressure, we obtain results in good agreement with available experimental and reproduce the type-I/type-II classification across all materials studied. The unified access to $\xi_0$ and $\lambda_\mathrm{L}$ further allows us to construct the Uemura plot entirely from first principles, 
showing that higher-$T_\mathrm{c}$ systems are characterized by the simultaneous realization of strong pairing and large phase stiffness. Our results establish a predictive first-principles route to superconducting length scales and provide a microscopic interpretation of empirical correlations in superconductivity.
\end{abstract}


\maketitle

{\it Introduction---}
Establishing a first-principles description of the superconducting state and quantitatively predicting its physical properties have been central challenges in solid state physics~\cite{doi:10.7566/JPSJ.87.041012,FLORESLIVAS20201,Boeri_2022,Sanna2024,PhysRevB.87.024505,PhysRevB.108.064511,Lucrezi2024}. Beyond the evaluation of the superconducting transition temperature ($T_\mathrm{c}$), a microscopic and parameter-free determination of quantities characterizing the superconducting state itself is of fundamental importance, not only for deepening our understanding of superconductivity but also for guiding materials design~\cite{https://doi.org/10.1002/adfm.202404043,Sanna2026} and technological applications~\cite{Dobrovolskiy_2026}.

In this context, superconducting density functional theory (SCDFT)~\cite{PhysRevLett.60.2430,PhysRevLett.86.2984,sanna2021introduction}, which extends density functional theory to the superconducting state, has emerged as a powerful theoretical framework. Since its explicit formulation in 2005, SCDFT has been successfully applied mainly to conventional phonon-mediated superconductors~\cite{PhysRevB.72.024545,PhysRevB.72.024546,PhysRevLett.94.037004,PhysRevLett.96.047003,PhysRevB.106.L180501}, enabling quantitative first-principles calculations of $T_\mathrm{c}$. More recently, systematic improvements of the exchange-correlation functional~\cite{PhysRevLett.125.057001}, including the effects of plasmons~\cite{PhysRevLett.111.057006,PhysRevB.102.214508} and spin fluctuations~\cite{PhysRevB.90.214504,PhysRevB.94.014503, PhysRevB.101.134511}, have significantly broadened the range of materials accessible within this framework~\cite{Boeri_2022,Sanna2024}.

Despite these advances, a first-principles evaluation of fundamental length scales characterizing the superconducting state---namely, the coherence length ($\xi_0$) and the magnetic penetration depth ($\lambda_\mathrm{L}$)---has not yet been established within SCDFT. These quantities play a pivotal role in classifying superconductors into type-I and type-II regimes, and directly determine critical magnetic fields, depairing currents, and superconducting stiffness. Therefore, these length scales are indispensable for both fundamental characterization and practical assessment of superconducting materials~\cite{Tinkham1996}.



A microscopic approach for evaluating the coherence length and penetration depth on the same footing was previously introduced within dynamical mean-field theory (DMFT)~\cite{Witt2024}.
In that work, superconducting states with finite center-of-mass momentum ($\mathbf{Q}$) were used to extract $\xi_0$ and $\lambda_\mathrm{L}$ from the momentum dependence of the order parameter and the supercurrent. Building on this finite-momentum framework, we develop a superconducting density functional theory (SCDFT) formulation providing a parameter-free first-principles methodology for evaluating superconducting length scales. Since these length scales are typically much larger than the lattice spacing, their determination hinges on resolving the superconducting response in the asymptotic small-$|\textbf{Q}|$ limit. We therefore derive the SCDFT gap equation with finite pairing momentum and develop computational techniques that ensure numerical stability in this regime, including the calculation of superconducting supercurrents.

We apply the 
resulting formalism to representative elemental superconductors, including Al, Nb, and Pb, and perform detailed comparisons with available experimental data. For all these materials, we find excellent agreement for both $\xi_0$ and $\lambda_\mathrm{L}$, demonstrating the quantitative reliability of our approach. We further extend our analysis to the A15~\cite{dmwh-1sbf} superconductor V$_3$Si and to H$_3$S, a high-temperature superconductor stabilized under high pressure~\cite{Drozdov2015,FLORESLIVAS20201}, and again obtain results in very good agreement with experimental estimates. Beyond this agreement, the unified determination of the length scales allows us to classify each material as type-I or type-II from first principles and to construct the Uemura plot without empirical input, placing the relation between pairing strength and phase stiffness on a microscopic basis.

In particular, H$_3$S becomes superconducting only above pressures of approximately 100~GPa~\cite{Drozdov2015,FLORESLIVAS20201}, making experimental determinations of $\xi_0$ and $\lambda_\mathrm{L}$ extremely challenging. In such cases, the first-principles SCDFT-based simulations developed in this work provide a powerful and indispensable tool for characterizing the superconducting state and predicting its properties under extreme conditions.

{\it Method---}
The superconducting singlet order parameter is computed as the expectation value of the product of annihilation operators, $\hat{\psi}_\sigma(\mathbf{r})$, as~\cite{PhysRevLett.60.2430}:
\begin{align}
  \chi(\mathbf{r},\mathbf{r}') &\equiv
  \left\langle{\hat{\psi}}_\uparrow(\mathbf{r}){\hat{\psi}}_\downarrow(\mathbf{r}')\right\rangle
  \nonumber \\
  &=\sum_{n\mathbf{k}}\{
    u_{n\mathbf{k}}(\mathbf{r})v_{n\mathbf{k}}^\ast(\mathbf{r}')f\left(-E_{n\mathbf{k}}\right)
  \nonumber \\
   &\qquad -u_{n\mathbf{k}}(\mathbf{r}')v_{n\mathbf{k}}^\ast(\mathbf{r})f(E_{n\mathbf{k}})\},
\end{align}
where $u_{n\mathbf{k}}(\mathbf{r})$, $v_{n\mathbf{k}}(\mathbf{r})$, and $E_{n\mathbf{k}}$ are the eigenvectors and eigenvalues of the Kohn--Sham--Bogoliubov--de Gennes eq.
\begin{align}
  \left(\begin{matrix}
    -\frac{\boldsymbol{\nabla}^2}{2}+V_\mathrm{KS}(\mathbf{r})-\mu & {\hat{\Delta}}_\mathrm{KS} \\
    {\hat{\Delta}}_\mathrm{KS}^\dag & \frac{\boldsymbol{\nabla}^2}{2}-V_\mathrm{KS}(\mathbf{r})+\mu
  \end{matrix}\right)&
  \left(\begin{matrix}
    u_{n\mathbf{k}}(\mathbf{r}) \\
    v_{n\mathbf{k}}(\mathbf{r})
  \end{matrix}\right)
\nonumber \\
  =E_{n\mathbf{k}}&
  \left(\begin{matrix}
    u_{n\mathbf{k}}(\mathbf{r}) \\
    v_{n\mathbf{k}}(\mathbf{r})
  \end{matrix}\right).
\end{align}
Next, we consider $\chi(\mathbf{r},\mathbf{r}')$ which is periodically modulated with a finite momentum $\mathbf{Q}$~\cite{PhysRev.135.A550}~(see Supplemental Material of Ref.~\cite{Witt2024}), 
\begin{eqnarray}
  \chi(\mathbf{r}+\mathbf{R},\mathbf{r}'+\mathbf{R})=e^{i\mathbf{Q}\cdot\mathbf{R}}\chi(\mathbf{r},\mathbf{r}').
\end{eqnarray}
This assumption leads to the decoupling approximation~\cite{PhysRevB.72.024545} together with the generalized Bloch theorem~\cite{sandratskii1998noncollinear,PhysRevB.96.195133,Witt2024},
\begin{align}
  v_{n\mathbf{k}}(\mathbf{r})&\approx v_{n\mathbf{k}}\varphi_{n\mathbf{k}-\mathbf{Q}/2}(\mathbf{r}),
  \\
  u_{n\mathbf{k}}(\mathbf{r})&\approx u_{n\mathbf{k}}\varphi_{n\mathbf{k}+\mathbf{Q}/2}(\mathbf{r}),
\end{align}
where $\varphi_{n\mathbf{k}}(\mathbf{r})$ is the normal-state Kohn--Sham orbital at band $n$ and wavenumber $\mathbf{k}$.
We obtain the scalar coefficients $u_{n\mathbf{k}}$ and $v_{n\mathbf{k}}$ by solving the $2\times2$ secular equation
\begin{eqnarray}
  \left(\begin{matrix}
    \xi_{n\mathbf{k}+\mathbf{Q}/2} & \Delta^{(\mathbf{Q})}_{n\mathbf{k}} \\
    \Delta^{(\mathbf{Q})*}_{n\mathbf{k}} & -\xi_{n\mathbf{k}-\mathbf{Q}/2} \\
  \end{matrix}\right)
  \left(\begin{matrix}u_{n\mathbf{k}} \\ v_{n\mathbf{k}}\\\end{matrix}\right)
  =E_{n\mathbf{k}}
  \left(\begin{matrix}u_{n\mathbf{k}}\\v_{n\mathbf{k}}\\\end{matrix}\right),
\end{eqnarray}
where $\xi_{n\mathbf{k}}$ is the Kohn--Sham eigenvalue measured relative to the Fermi level, and the gap function $\Delta^{(\mathbf{Q})}_{n\mathbf{k}}$ fulfills following Kohn--Sham gap equation,
\begin{align}
  &\Delta_{n\mathbf{k}}^{(\mathbf{Q})}=
-Z^{(\mathbf{Q})}_{n\mathbf{k}} \Delta_{n\mathbf{k}}^{(\mathbf{Q})}
  -\frac{1}{4}\sum_{n'\mathbf{k}'}K^{(\mathbf{Q})}_{n\mathbf{k}n'\mathbf{k}'}
  \frac{\Delta_{n'\mathbf{k}'}^{(\mathbf{Q})}}{E_{n'\mathbf{k}'}^{(\mathbf{Q})}}
\nonumber \\
  &\times
  \left\{
    \tanh\left(\frac{E_{n'\mathbf{k}'}^{(\mathbf{Q})}+d_{n'\mathbf{k}'}^{(\mathbf{Q})}}{2T}\right)
   +\tanh\left(\frac{E_{n'\mathbf{k}'}^{(\mathbf{Q})}-d_{n'\mathbf{k}'}^{(\mathbf{Q})}}{2T}\right)
  \right\},
\label{eq:gap_org}
\end{align}
where $E_{n\mathbf{k}}^{(\mathbf{Q})}\equiv\sqrt{\bar{\xi}_{n\mathbf{k}}^{(\textbf{Q})2}+|\Delta_{n\mathbf{k}}^{(\mathbf{Q})}|^2}$, ${\bar{\xi}}_{n\mathbf{k}}^{(\mathbf{Q})}\equiv(\xi_{n\mathbf{k}+\mathbf{Q}/2}+\xi_{n\mathbf{k}-\mathbf{Q}/2})/2$, and $d_{n\mathbf{k}}^{(\mathbf{Q})}\equiv(\xi_{n\mathbf{k}+\mathbf{Q}/2}-\xi_{n\mathbf{k}-\mathbf{Q}/2})/2$.
The renormalization factor $Z^{(\mathbf{Q})}_{n\mathbf{k}}$ and the kernel $K^{(\mathbf{Q})}_{n\mathbf{k}n'\mathbf{k}'}$ are computed with the functional derivative of the exchange-correlation grandpotentaial involving normal- ($\Omega^\textrm{(n)}_\textrm{xc}$) and anomalous- ($\Omega^\textrm{(a)}_\textrm{xc}$) Kohn--Sham Green's function as follows:
\begin{align}
 Z^{(\textbf{Q})}_{n\mathbf{k}} &\equiv
\left. \left(
\frac{1}{\Delta^{(\textbf{Q})}_{n\textbf{k}}}\frac{\delta \Omega^\textrm{(n)}_\mathrm{xc}}{\delta \chi^{\textbf{Q}*}_{n\textbf{k}}}
\right)\right|_{\{\Delta_{n\textbf{k}}\}=0},
\\
K^{(\textbf{Q})}_{n\mathbf{k}n'\mathbf{k}'}&\equiv 
\left. 
\frac{\delta^2
\Omega^\textrm{(a)}_\mathrm{xc} }
{\delta \chi^{\textbf{Q}*}_{n\textbf{k}}\delta \chi^{(\textbf{Q})}_{n'\textbf{k}'}}
\right|_{\{\Delta_{n\textbf{k}}\}=0},
\\
\chi^{(\mathbf{Q})}_{n\mathbf{k}} &\equiv \iint d^3r d^3 r' \varphi^*_{n\mathbf{k}+\textbf{Q}/2}(\mathbf{r})
\chi(\mathbf{r},\mathbf{r}') \varphi_{n\mathbf{k}-\textbf{Q}/2}(\mathbf{r}').
\end{align}
$\Omega^\textrm{(n/a)}_\mathrm{xc}$ contains the contribution from the electron-phonon coupling~\cite{PhysRevLett.125.057001}, dynamically screened Coulomb interaction~\cite{PhysRevLett.111.057006,PhysRevB.102.214508}, and the spin-fluctuation~\cite{PhysRevB.90.214504,PhysRevB.102.214515}.
To stabilize the numerical $\mathbf{k}$-point integration, we introduce an auxiliary energy-dependent gap function~\cite{PhysRevB.95.054506},  which satisfies $\Delta^{(\mathbf{Q})}_{n\mathbf{k}} \equiv \Delta^{(\mathbf{Q})}_{n\mathbf{k}}(0)$, and
\begin{align}
  \Delta_{n\mathbf{k}}^{(\mathbf{Q})}(\xi)&=
  -\frac{1}{2}\int{d\xi'} \sum_{n'\mathbf{k}'} \delta(\xi'-{\bar{\xi}}_{n'\mathbf{k}'}^{(\mathbf{Q})})
  \nonumber \\
   &\times
   \frac{K_{n\mathbf{k}n'\mathbf{k}'}(\xi,\xi')}{1+Z_{n\mathbf{k}}(\xi)}
   \frac{\Delta_{n'\mathbf{k}'}^{(\mathbf{Q})}(\xi')}{E_{n'\mathbf{k}'}^{(\mathbf{Q})}(\xi')}
   \tanh{\left(\frac{E_{n'\mathbf{k}'}^{(\mathbf{Q})}(\xi')}{2T}\right)}
   \nonumber \\
  &\times\frac{1-\tanh^2{\left(\frac{d_{n'\mathbf{k}'}^{(\mathbf{Q})}(\xi')}{2T}\right)}}
     {1-\tanh^2{\left(\frac{E_{n'\mathbf{k}'}^{(\mathbf{Q})}(\xi')}{2T}\right)}\tanh^2{\left(\frac{d_{n'\mathbf{k}'}^{(\mathbf{Q})}(\xi')}{2T}\right)}},
\label{eq:gap}
\end{align}
where $E_{n\mathbf{k}}^{(\mathbf{Q})}(\xi)\equiv\sqrt{\xi^2 + |\Delta_{n\mathbf{k}}^{(\mathbf{Q})}(\xi)|}$, and
\begin{align}
  d_{n\mathbf{k}}^{(\mathbf{Q})}(\xi)\equiv
  \frac{\left\langle
    \left(\xi_{n\mathbf{k}+\mathbf{Q}/2}-\xi_{n\mathbf{k}-\mathbf{Q}/2}\right)\delta(\xi_{n\mathbf{k}}-\xi)
  \right\rangle}
  {2\left\langle \delta\left(\xi_{n\mathbf{k}}-\xi\right) \right\rangle}
\end{align}
is computed as an iso-energy average with the $\mathbf{k}$-dependent density of states as a weight.
In Eq.~(\ref{eq:gap}), we ignore the $\mathbf{Q}$ dependence of $K_{n\mathbf{k}n'\mathbf{k}'}$ and $Z_{n\mathbf{k}}$.
In theoretical treatments of superconducting states with a finite center-of-mass momentum, it is common to neglect the explicit $\mathbf{Q}$ dependence of the pairing and mass-renormalization kernels, while retaining the $\mathbf{Q}$ dependence of the single-particle energies and anomalous density.
This approximation is routinely adopted in model calculations because the relevant center-of-mass momentum is typically of the order of the inverse coherence length, $|\textbf{Q}|\sim\xi_0^{-1}$, and is therefore much smaller than the characteristic microscopic momentum scales governing the interaction kernels.
Additionally, in the Supplemental Material, we demonstrate that, as a representative component of the kernel, the static Coulomb contribution is $\textbf{Q}$-independent in two limiting cases: the nearly free-electron model and the strictly localized tight-binding model.
For materials with a coherence length much larger than the lattice constant (such as Al and Nb), we need to compute at tiny $\mathbf{Q}$ relative to the $\mathbf{k}$-grid spacing.
Therefore, the Kohn--Sham energy at the $\mathbf{k}\pm\mathbf{Q}/2$ point, which is slightly shifted from the original grid point, is evaluated using the Taylor expansion as
\begin{align}
  \xi_{n\mathbf{k}+\mathbf{Q}/2} &= \xi_{n\mathbf{k}}
  +\sum_{\alpha=x,y,z}{\frac{\partial\xi_{n\mathbf{k}}}{{\partial k}_\alpha}\frac{Q_\alpha}{2}}
  \nonumber \\
  &+\frac{1}{2}
  \sum_{\alpha,\alpha'=x,y,z}{\frac{\partial^2\xi_{n\mathbf{k}}}{{\partial k}_\alpha{\partial k}_{\alpha'}}\frac{Q_\alpha}{2}\frac{Q_{\alpha'}}{2}}.
\end{align}

To compute the penetration depth, we need the spatial average of the supercurrent density
\begin{align}
  \bar{\mathbf{j}}_\textrm{sc}^{(\mathbf{Q})} &\equiv
  \bar{\mathbf{j}}^{(\mathbf{Q})} - 
  \bar{\mathbf{j}}^{(\mathbf{Q})}\left(\Delta=0\right),
  \\
  \bar{\mathbf{j}}^{(\mathbf{Q})}&\equiv
  \frac{1}{V}\int dr^3 \mathbf{j}^{(\mathbf{Q})}(\mathbf{r}),
\end{align}
where $\mathbf{j}^{(\mathbf{Q})}(\mathbf{r})$ is the expectation value of the current density,
\begin{align}
  \mathbf{j}(\mathbf{r}) &=
  \frac{i}{2}\sum_{\sigma}\left\langle\hat{\psi}_\sigma^\dag(\mathbf{r})\boldsymbol{\nabla}\hat{\psi}_\sigma(\mathbf{r})-\hat{\psi}_\sigma(\mathbf{r})\boldsymbol{\nabla}\hat{\psi}_\sigma^\dag(\mathbf{r})\right\rangle
\end{align}
for a finite-momentum superconducting state.
With the same approximation as derived above, we obtain
\begin{align}
  \bar{\mathbf{j}}_\textrm{sc}^{(\mathbf{Q})}
  &=\frac{1}{V}\int{d\xi}\sum_{n\mathbf{k}}
  \delta(\xi-{\bar{\xi}}_{n\mathbf{k}}^{(\mathbf{Q})})
  \mathbf{v}_{n\mathbf{k}+\mathbf{Q}/2}
  \nonumber \\
  &\times\left[
    \frac{1}{2}
    \left\{
      \tanh\left(\frac{d_{n\mathbf{k}}^{\mathbf{Q}}(\xi)+E_{n\mathbf{k}}^{\mathbf{Q}}(\xi)}{2T}\right)
    \right.\right. \nonumber \\ &\left. \qquad
     +\tanh\left(\frac{d_{n\mathbf{k}}^{\mathbf{Q}}(\xi)-E_{n\mathbf{k}}^{\mathbf{Q}}(\xi)}{2T}\right)
    \right\}
    \nonumber \\
    & \quad
    +\frac{\xi}{2E_{n\mathbf{k}}^{\mathbf{Q}}(\xi)}
    \left\{
      \tanh\left(\frac{d_{n\mathbf{k}}^{\mathbf{Q}}(\xi)+E_{n\mathbf{k}}^{\mathbf{Q}}(\xi)}{2T}\right)
    \right. \nonumber \\ & \left. \qquad\qquad\qquad
    -\tanh\left(\frac{d_{n\mathbf{k}}^{\mathbf{Q}}(\xi)-E_{n\mathbf{k}}^{\mathbf{Q}}(\xi)}{2T}\right)
    \right\}
    \nonumber \\
    &\quad \left.-\tanh\left(\frac{d_{n\mathbf{k}}^{\mathbf{Q}}(\xi)+\xi}{2T}\right)
  \right],
\label{eq:current}
\end{align}
where $\mathbf{v}_{n\mathbf{k}}$ is the Fermi velocity.
Detailed derivation can be seen in the Supplemental Material~\cite{SM}.
\begin{table*}[!htb]
\begin{center}
\begin{tabular}{ccccccccc}
  \hline
material & Al & In & Sn & Ta & Pb & Nb & H$_3$S & V$_3$Si \\
  \hline
\rowcolor{gray!10}
 Exp. $T_\mathrm{c}$ (K) & 1.140~\cite{kittel2004introduction} & 3.4035~\cite{kittel2004introduction} & 3.722~\cite{kittel2004introduction} & 4.483~\cite{kittel2004introduction} & 7.193~\cite{kittel2004introduction} & 9.50~\cite{kittel2004introduction} & 184~\cite{Drozdov2015} & 17.1~\cite{kittel2004introduction} \\
 Calc. $T_\mathrm{c}$ (K)  & 1.61 & 3.06 & 4.02 & 4.42 & 5.33 & 8.09 & 181 &  20.5 \\
\rowcolor{gray!10}
Exp. $\xi_0$ (nm) & \begin{tabular}{c}500--1,600~\cite{buckel1991superconductivity}\\1,600~\cite{HUEBENER1990605,orlando1991foundations}\end{tabular} & \begin{tabular}{c}260~\cite{PhysRev.123.442}\\275~\cite{HUEBENER1990605}\\360~\cite{orlando1991foundations}\\360--440~\cite{buckel1991superconductivity}\\440~\cite{van1981principles}\end{tabular} & \begin{tabular}{c}94~\cite{HUEBENER1990605}\\120--320~\cite{buckel1991superconductivity}\\230~\cite{orlando1991foundations,van1981principles}\end{tabular} & \begin{tabular}{c}92.5(4)~\cite{PhysRevLett.22.229}\\92.5~\cite{PhysRevB.7.136}\\93~\cite{buckel1991superconductivity}\end{tabular} & \begin{tabular}{c}51--83~\cite{buckel1991superconductivity}\\74~\cite{HUEBENER1990605}\\83~\cite{kittel2004introduction,van1981principles}\\90(5)~\cite{PhysRevB.72.024506}\\90~\cite{orlando1991foundations}\end{tabular} & \begin{tabular}{c}38~\cite{kittel2004introduction,van1981principles}\\39--40~\cite{buckel1991superconductivity}\\39.9(25)~\cite{10.1103/2nsw-n8gf}\\40~\cite{orlando1991foundations}\end{tabular} & 2~\cite{Mozaffari2019} & \begin{tabular}{c}3~\cite{orlando1991foundations}\\4~\cite{Zehetmayer_2014,buckel1991superconductivity}\end{tabular} \\
Calc. $\xi_0$ (nm) & 636 & 254 & 132 & 80 & 104 & 34 & 3.0 & 2.2 \\
 $\xi_\textrm{Pippard,ep}$ (nm)  & 712 & 264 & 168 & 82.2 & 89.7 & 34 & 1.46 & 1.74 \\
\rowcolor{gray!10}
Exp. $\lambda_\textrm{L}$ (nm) & \begin{tabular}{c}15.7~\cite{Lopez-Nunez_2025}\\16~\cite{kittel2004introduction,orlando1991foundations}\\50~\cite{buckel1991superconductivity,HUEBENER1990605}\end{tabular} & \begin{tabular}{c}21~\cite{van1981principles}\\24--64~\cite{buckel1991superconductivity}\\29.9~\cite{Raychaudhuri1985}\\47~\cite{HUEBENER1990605}\\65~\cite{orlando1991foundations}\end{tabular} & \begin{tabular}{c}25--50~\cite{buckel1991superconductivity}\\34~\cite{kittel2004introduction}\\36~\cite{van1981principles}\\50~\cite{orlando1991foundations}\\52~\cite{HUEBENER1990605}\end{tabular} & \begin{tabular}{c}33(1)~\cite{PhysRevLett.22.229}\\34.5~\cite{PhysRevB.7.136}\\35~\cite{buckel1991superconductivity}\end{tabular} & \begin{tabular}{c}32--39~\cite{buckel1991superconductivity}\\37~\cite{kittel2004introduction,van1981principles}\\40~\cite{orlando1991foundations}\\47~\cite{HUEBENER1990605}\end{tabular} & \begin{tabular}{c}27(3)~\cite{PhysRevB.72.024506}\\29.1(10)~\cite{10.1103/2nsw-n8gf}\\32--44~\cite{buckel1991superconductivity}\\39~\cite{kittel2004introduction,van1981principles}\\85~\cite{orlando1991foundations}\end{tabular} & \begin{tabular}{c}20~\cite{Minkov2022}\\37~\cite{Minkov2023}\end{tabular} & \begin{tabular}{c}60~\cite{orlando1991foundations}\\70~\cite{buckel1991superconductivity}\\90~\cite{Zehetmayer_2014}\end{tabular} \\
Calc. $\lambda_\textrm{L}$ from peak (nm) & 26 & 33 & 43 & 34 & 24 & 40 & 22 & 136 \\
Calc. $\lambda_\textrm{L}$ from slope (nm) & 21 & 28 & 43 & 33 & 22 & 34 & 19 & 97 \\
\rowcolor{gray!10}
Exp. $\lambda_{\mathrm{L}}/\xi_0$ & 0.0098--0.1 & 0.05--0.25 & 0.11--0.55 & 0.35--0.38 & 0.36--0.92 & 0.68--2.24 & 10--19 & 15--30 \\
 Calc. $\lambda_{\mathrm{L}}/\xi_0$ & 0.033--0.041 & 0.11--0.13  & 0.33 &  0.41--0.43 & 0.21--0.23 & 1.0--1.2 & 6.3--7.3 & 44--62 \\
\hline
\end{tabular}
\end{center}
\caption{Calculated and experimental values for the transition temperature ($T_\mathrm{c}$), coherence length ($\xi_0$), penetration depth ($\lambda_\textrm{L}$), and the Ginzburg--Landau parameter ($\kappa=\lambda_{\mathrm{L}}/\xi_0$) for elemental bulk materials and compounds. The vertical line separates type I (left) from type II (rigth) superconductors.
Both $\lambda_\textrm{L}$ values, calculated from the peak [Eq.~(\ref{eq_lambda_peak})] and slope [Eq.~(\ref{eq_lambda_slope})] of $\bar{\mathbf{j}}_{\mathrm{sc}}^{(Q)}$, are shown.
To compare the experimental results for type-I superconductors, we also show the Pippard coherence length including the electron-phonon renormalization $\xi_\textrm{Pippard,ep}$.
The data in Refs.~\onlinecite{HUEBENER1990605,orlando1991foundations,van1981principles} were obtained from Ref.~\onlinecite{poole2014superconductivity}.
}
\label{tab:results}
\end{table*}

{\it Results---}
We apply the methodology outlined above to Nb, a prototypical conventional superconductor. As an initial benchmark, we perform the standard SCDFT calculation for Cooper pairs with ${\mathbf Q}=\mathbf{0}$ and evaluate $T_\mathrm{c}$ using Superconducting-Toolkit~\cite{PhysRevB.101.134511}. The calculated transition temperature is 8.09 K, in good agreement with the experimental value of 9.50 K. This level of agreement demonstrates the reliability of the numerical framework adopted in the present study.

\begin{figure}[tbp]
\includegraphics[width=1\columnwidth]{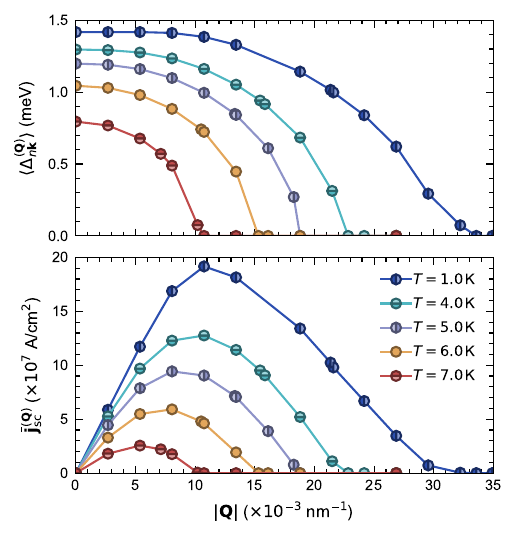}
\caption{(Top) Fermi-surface-averaged superconducting gap function $\langle\Delta_{n\mathbf{k}}^{(\mathbf{Q})}\rangle$ plotted as a function of the momentum $|\mathbf{Q}|$ of the Cooper pair.
(Bottom) Supercurrent density $\bar{\mathbf{j}}_{\mathrm{sc}}^{(Q)}$ as a function of $|\mathbf{Q}|$.}
\label{fig:deltaq_nb}
\end{figure}

Next, we calculate the gap function $\Delta_{n\mathbf{k}}^{(\mathbf{Q})}$ for Cooper pairs with finite ${\mathbf Q}$ following Eq.~(\ref{eq:gap}). Figure~\ref{fig:deltaq_nb} shows the momentum dependence of the Fermi-surface-averaged gap function $\langle\Delta_{n\mathbf{k}}^{(\mathbf{Q})}\rangle$ for temperatures ranging from 1~K to 7~K. Finite pairing momenta generally suppress the gap, and at temperatures close to $T_{\mathrm{c}}$, $\Delta_{n\mathbf{k}}^{(\mathbf{Q})}$ 
follows the Ginzburg--Landau expression
\begin{align}
    \langle\Delta_{n\mathbf{k}}^{(\mathbf{Q})}\rangle=\langle\Delta_{n\mathbf{k}}^{(\mathbf{Q=0})}\rangle\sqrt{1-\xi_0^2Q^2},
	\label{eq:q_dependence_gap}
\end{align}
where $Q=|\mathbf Q|$. Following Ref.~\cite{Witt2024}, we define $Q_2$ as the wave number at which $\langle\Delta_{n\mathbf{k}}^{(\mathbf{Q})}\rangle$ is reduced to $1/\sqrt{2}$ of its zero-momentum value $\langle\Delta_{n\mathbf{k}}^{(\mathbf{Q=0})}\rangle$, and evaluate the coherence length as $\xi_0 = 1/(\sqrt{2} Q_2)$. Although Eq.~(\ref{eq:q_dependence_gap}) is strictly valid only close to $T_{\mathrm{c}}$, the $Q_2$ criterion provides a reasonable estimate even at lower temperatures. Indeed, applying this approach, we extract a zero-temperature coherence length $\xi_0 = 34$~nm which is in very good agreement with the experimental estimate of $39 \pm 1$~nm~\cite{10.1103/2nsw-n8gf,kittel2004introduction,van1981principles,buckel1991superconductivity,orlando1991foundations}.

We also evaluate the Pippard coherence length 
\begin{align}
\xi_\textrm{Pippard} \equiv \frac{\left\langle|\mathbf{v}_{n\textbf{k}}|\right\rangle}{\left\langle\Delta^{(\mathbf{0})}_{n\textbf{k}}\right\rangle\pi},
\end{align}
where $\langle |\mathbf{v}_{n\mathbf{k}}| \rangle$ is the Fermi-surface-averaged Fermi velocity.
This procedure yields a value of 79.6~nm. Taking into account the electron--phonon coupling constant $\lambda_{\rm ep}=1.34$ for Nb, the Fermi velocity is effectively renormalized by a factor of $(1+\lambda_{\rm ep})$, and the Pippard coherence length is reduced as $\xi_\textrm{Pippard,ep}\equiv\xi_\textrm{Pippard}(1+\lambda_{\rm ep})^{-1}$. Finally, $\xi_\textrm{Pippard,ep}$ becomes 34~nm, which is consistent with the value of $\xi_0$ obtained above.

We next evaluate the magnetic penetration depth. Using Eq.~(\ref{eq:current}), we calculate the supercurrent density $\bar{\mathbf{j}}_{\rm sc}^{(\mathbf{Q})}$ carried by Cooper pairs with finite $\mathbf{Q}$. The results are shown in Fig.~\ref{fig:deltaq_nb}. The current density $\bar{\mathbf{j}}_{\rm sc}^{(\mathbf{Q})}$ increases linearly for small $\mathbf{Q}$, reaches a maximum, and then decreases as $\mathbf{Q}$ increases further. From the linear-response slope of $\bar{\mathbf{j}}_{\rm sc}^{(\mathbf{Q})}$ 
at $\mathbf{Q}=\mathbf{0}$, the magnetic penetration depth is evaluated as
\begin{equation}
\lambda_\mathrm{L}
=
\left.
\left(
2\mu_0
\frac{\partial \bar{\mathbf j}_{\mathrm{sc}}^{(Q)}}{\partial Q}
\right)^{-1/2}
\right|_{Q=0},
\label{eq_lambda_slope}
\end{equation}
where $\mu_0$ denotes the vacuum permeability. This procedure yields a zero-temperature penetration depth of $\lambda_\mathrm{L} = 34$~nm in good agreement with the experimental range of 27--39~nm~\cite{PhysRevB.72.024506,10.1103/2nsw-n8gf,kittel2004introduction,van1981principles,buckel1991superconductivity,orlando1991foundations}.

As an independent estimate, $\lambda_\mathrm{L}$ can also be inferred within Ginzburg--Landau theory by combining the depairing current $J_{\mathrm{dp}}$, i.e., the maximum value of $\bar{\mathbf{j}}_{\mathrm{sc}}^{(\mathbf{Q})}$, with the zero-temperature coherence length $\xi_0$,
\begin{equation}
\lambda_\mathrm{L} = \sqrt{\frac{\Phi_0}{3\sqrt{3}\mu_0 \xi_0 J_{\rm dp}}},
\label{eq_lambda_peak}
\end{equation}
where $\Phi_0$ is the flux quantum.
This approach yields $\lambda_\mathrm{L} = 40$~nm, which remains in good agreement with experiment. While Eq.~(\ref{eq_lambda_peak}) relies on the Ginzburg--Landau approximation and therefore provides an approximate estimate, the overall agreement demonstrates the robustness of the present approach.

We note that a particular strength of our framework is that $J_{\mathrm{dp}}$ is directly obtained from $\bar{\mathbf{j}}_{\rm sc}^{(\mathbf{Q})}$ without any further assumption. The depairing current constitutes the intrinsic upper bound for dissipationless current transport, whereas experimentally measured critical currents $J_{\mathrm{c}}$ are typically reduced by vortex motion, sample geometry, grain boundaries, and other forms of disorder or inhomogeneity~\cite{Tinkham1996,RevModPhys.34.667,Ruiz2026}. This distinction is particularly relevant in type-II superconductors, where vortex dynamics generally suppress $J_{\mathrm{c}}$ well below $J_{\mathrm{dp}}$~\cite{Blatter1994}. For Nb, we obtain $J_{\mathrm{dp}}=19\times10^7$\,A/cm$^2$ which is about 25 times larger than measured $J_{\mathrm{c}}=0.75\times10^7$\,A/cm$^2$ (see Table S2 in the Supplemental Material~\cite{SM}).

We have applied the same procedure to the set of superconductors listed in Table~\ref{tab:results}. The most demanding cases are low-$T_\mathrm{\mathrm{c}}$ superconductors such as Al, In and Sn, whose coherence lengths are exceptionally long and require resolving the suppression of the order parameter at extremely small momenta $\mathbf{Q}$, corresponding to less than $\sim0.1\,\%$ of the Brillouin zone. The corresponding Pippard lengths (Table~\ref{tab:results}) provide an additional point of comparison less sensitive to this limitation. Benchmarking against the available experimental literature highlights the substantial spread of reported coherence lengths and penetration depths, particularly for type-I superconductors. The reported values depend on the experimental probe and the procedure used to extract these quantities, illustrating that a unique experimental determination is often difficult. This underscores the importance of predictive first-principles calculations, which provide direct access to intrinsic superconducting length scales and thereby complement experimental characterization. The calculated coherence lengths and magnetic penetration depths show good overall agreement with experiment. In particular, the correct classification of each superconductor as type-I ($\lambda_{\mathrm{L}}/\xi_0<1/\sqrt{2}$) or type-II ($\lambda_{\mathrm{L}}/\xi_0>1/\sqrt{2}$) is quantitatively reproduced.


Let us here emphasize that first-principles calculations enable the prediction of superconducting properties for materials under extreme conditions where experiments are challenging. Table~\ref{tab:results} includes results for H$_3$S, a high-temperature superconductor stabilized under high pressure~\footnote{As a representative case for H$_3$S, we performed simulations at 200 GPa with the $Im\bar{3}m$ structure, which is expected from both experiments~\cite{Einaga2016} and theory~\cite{Duan2014,PhysRevLett.114.157004,PhysRevB.91.224513,Errea2016,Flores-Livas2016,PhysRevLett.117.075503}. Our calculations quantitatively reproduced $T_\mathrm{c}$ (Table \ref{tab:results}). Although anharmonicity, which affects structural stability~\cite{Errea2016} and $T_\mathrm{c}$~\cite{PhysRevLett.114.157004,PhysRevB.93.094525}, was not considered in this study, our method can be straightforwardly integrated with anharmonic phonon theories~\cite{PhysRevB.89.064302,PhysRevB.92.054301}.}.
Our calculations yield $\xi_0 = 3.0$~nm and $\lambda_\mathrm{L} = 19$--22~nm, indicating that H$_3$S is a type-II superconductor. Based on high-field measurements, the upper critical field $H_{\rm c2}$(T) of H$_3$S is estimated, using the Werthamer--Helfand--Hohenberg theory~\cite{PhysRev.147.295}, to extrapolate to approximately 70-90 T as $T\rightarrow 0$~\cite{Mozaffari2019}.
Using the relation $H_{\rm c2} = \Phi_0/(2\pi \xi_0^2)$, this value corresponds to $\xi_0 \simeq 2$~nm, in good agreement with our calculated result. Experimental estimates of $\lambda_\mathrm{L}$ in H$_3$S have been reported by Minkov \textit{et al}.~in two independent studies using SQUID magnetometry under megabar pressures.
From magnetization curves, $\lambda_\mathrm{L}$ was estimated to be about 20\,nm~\cite{Minkov2022}, while a later study based on trapped-flux measurements yielded a somewhat larger value of about 37\,nm at low temperature~\cite{Minkov2023}. The difference reflects the sensitivity of the measurement protocol and sample-geometry corrections under extreme pressure conditions. Our theoretical value ($\sim20$\,nm) lies within this experimentally inferred range. Notably, we find an extremely large depairing current of $J_{\mathrm{dp}}=697\times10^7$\,A/cm$^2$ for H$_3$S, indicating an exceptional intrinsic current-carrying potential that substantially exceeds experimentally reported critical current scales in cuprate superconductors
~\cite{Stangl2021,Yao2021,Semwal2004}. Since this quantity is not experimentally accessible at the moment, our calculations provide a theoretical prediction that positions H$_3$S as a superconductor capable of sustaining an exceptionally large supercurrent.

\begin{figure}[tbp]
\includegraphics[width=1\columnwidth]{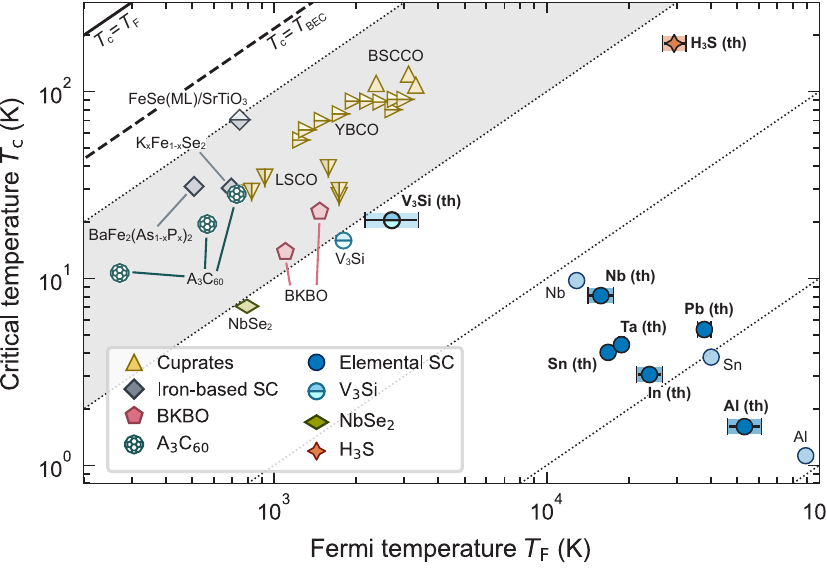}
\caption{\label{fig:uemura} Uemura plot: Log-log plot of the superconducting critical temperature $T_{\mathrm{c}}$ versus the Fermi temperature  $T_{\mathrm{F}}$. \textit{Ab initio} SCDFT results are highlighted by saturated symbols, while experimental values for comparison across a wide range of superconducting materials are shown in lighter tones~\cite{WittPhDThesis2024,Nakagawa2021,Uemura2019}. The solid line corresponds to $T_{\mathrm{c}} = T_{\mathrm{F}}$,
the dashed line indicates the critical temperature of a non-interacting BEC in 3D with
$T_{\mathrm{BEC}} = T_{\mathrm{F}}/4.16$, and the dotted lines serve as visual guides for successive decades of the $T_{\mathrm{c}}/T_{\mathrm{F}}$ ratio. The gray shaded area highlights the region where most unconventional superconductors are located, spanning $T_{\mathrm{c}}/T_{\mathrm{F}}$ ratios between 0.1 and 0.01. Horizontal bars on the theoretical points represent the range of Fermi temperatures obtained from SCDFT.}
\end{figure}

A key feature of the present approach is that the three key quantities, $T_\mathrm{c}$, $\xi_0$, and $\lambda_\mathrm{L}$, can be evaluated consistently within a single first-principles framework. 
One intriguing outcome of this unified treatment is the ability to construct the so-called Uemura plot~\cite{Uemura1989,Uemura1991}, which compares $T_\mathrm{c}$ with the Fermi temperature $T_\mathrm{F}$. The latter characterizes the phase stiffness of the superconducting condensate and is in three dimensions given by
\begin{align}
T_\mathrm{F} = \frac{1}{2}(3\pi^2)^{2/3}\frac{n_\mathrm{s}^{2/3}}{m^*}
\end{align}
where $m^*\equiv m^*_\mathrm{band}(1+\lambda_\mathrm{ep})$ is the effective mass (containing the band-structure part $m^*_\mathrm{band}$ and electron-phonon mass-enhancement), the superfluid density $n_\mathrm{s}$ is related to $\lambda_\mathrm{L}$ as $n_\mathrm{s} = m^*c^2/(4\pi \lambda_\mathrm{L}^2)$, and $c$ is the speed of light.
Our first-principles calculations naturally reproduce the experimentally observed trend that elemental superconductors, which are well described by conventional pairing mechanisms, systematically exhibit small values of $T_\mathrm{c}/T_\mathrm{F}$ (Fig.~\ref{fig:uemura}). In contrast, superconductors with higher $T_\mathrm{c}$'s, such as V$_3$Si and H$_3$S, are characterized by the simultaneous realization of a short $\xi_0$, indicative of strong pairing, and a short $\lambda_\mathrm{L}$, reflecting a large phase stiffness. These results demonstrate that the Uemura plot, which has traditionally been discussed in an experimental or semi-phenomenological context, can be understood and predicted on a fully first-principles basis that explicitly connects the pairing scale and the phase stiffness of the superconducting state.

{\it Summary---}
We have developed a first-principles methodology to evaluate the coherence length and magnetic penetration depth of superconductors within SCDFT by explicitly incorporating finite-momentum Cooper pairs. This approach enables a consistent determination of $T_\mathrm{c}$, $\xi_0$, and $\lambda_\mathrm{L}$ without phenomenological input and reproduces experimental trends across a wide range of superconducting materials. By constructing the Uemura plot directly from first principles, we demonstrate that empirical correlations between transition temperature, pairing strength, and phase stiffness can be understood on a microscopic basis. Our work establishes superconducting length scales as predictive quantities in first-principles theory and provides a foundation for exploring superconductivity under extreme conditions where experiments are challenging.

\textit{Acknowledgments---} We are grateful to Yuji Nakagawa and Yoshihiro Iwasa for sharing the experimental data of the Uemura plot. N.W. thanks the University of Tokyo for hospitality during
his research stay, where parts of this work
were conceived. This work is supported by Grant-in-Aid for Scientific Research from JSPS, KAKENHI Grant  No. 25H01246, No. 25H01252, No. 24H00190, No. 25K07220, No. 25K07234, No. 25H00420, JST K-Program JPMJKP25Z7, RIKEN TRIP initiative (RIKEN Quantum, Advanced General Intelligence for Science Program, Many-body Electron Systems).
N.~W. acknowledges funding support
from the Deutsche Forschungsgemeinschaft (DFG, German Research Foundation) under the SFB 1170 ``Tocotronics'' (project No. 258499086)  and Germany's Excellence Strategy through
the W\"urzburg-Dresden Cluster of Excellence on Complexity, Topology and Dynamics in Quantum Matter---ctd.qmat (EXC2147, project No. 390858490).
The numerical calculation in this paper was carried out on the supercomputer in the Information Technology Center at the University of Tokyo.

{\it Data availability---}
The data supporting this study's findings are available on the author's Web page~\cite{dataweb}.
\bibliography{ref}

\end{document}



\title{First-principles calculation of coherence length and penetration depth based on density functional theory for superconductors\\
supplemental material}


\author{Mitsuaki Kawamura}
\email[]{kawamura-mitsuaki-zr@ynu.ac.jp}
\homepage[]{https://mkawamura.ynu.ac.jp/}
\affiliation{Division of Intelligent Systems Engineering, Yokohama National University, Yokohama 240-8501, Japan}

\author{Takuya Nomoto}
\affiliation{Department of Physics, Tokyo Metropolitan University, Hachioji 192-0397, Japan}

\author{Niklas Witt}

\affiliation{Institute for Theoretical Physics and Astrophysics and W\"urzburg-Dresden Cluster
 of Excellence ct.qmat, University of W\"urzburg, W\"urzburg 97074, Germany}

\author{Ryotaro Arita}
\affiliation{Department of Physics, University of Tokyo, Hongo, Tokyo 113-0033, Japan}
\affiliation{Center for Emergent Matter Science, RIKEN, Wako 351-0198, Japan}

\date{\today}

\begin{abstract}
\end{abstract}


\maketitle

\section{Method}

\subsection{Ginzburg--Landau Theory for a Finite-Momentum Order Parameter}

We start from the Ginzburg--Landau (GL) free-energy functional for the superconducting order parameter
$\Psi(\mathbf r)$:

\begin{align}
F_{\rm GL}[\Psi]
=\int d^3 r\left[
\alpha(T)|\Psi(\mathbf r)|^2+\frac{b}{2}|\Psi(\mathbf r)|^4
+\frac{1}{2m^\ast}\left|
\left(i\nabla+2\mathbf A\right)\Psi(\mathbf r)
\right|^2
\right],
\end{align}
where $\alpha(T)$ is the temperature-dependent quadratic coefficient,
$b>0$ is the quartic coefficient, $m^\ast$ is the effective mass,
and $\mathbf A$ denotes the vector potential.

For a superconducting state with Cooper pairs carrying a finite center-of-mass momentum $\mathbf{Q}$, the order parameter consists of a single plane-wave component:
\begin{align}
\Psi(\mathbf r)=\Psi_{\mathbf Q}e^{i\mathbf Q\cdot \mathbf r}.
\label{2}
\end{align}
Substituting Eq.~(\ref{2}) into the gradient term, we obtain
\begin{align}
\left(i\nabla+2\mathbf A\right)\Psi(\mathbf r)
=\left(-\mathbf Q+2\mathbf A\right)\Psi(\mathbf r),
\end{align}
and therefore,
\begin{align}
\left|\left(i\nabla+2\mathbf A\right)\Psi(\mathbf r)\right|^2
=|\Psi_{\mathbf Q}|^2
\left(\mathbf Q^2-4\mathbf Q\cdot\mathbf A+4\mathbf A^2\right).
\end{align}
Since $|\Psi(\mathbf r)|^2=|\Psi_{\mathbf Q}|^2$ is spatially uniform, the free-energy density becomes

\begin{align}
\frac{F_{\rm GL}[\Psi]}{V}
=\alpha(T)|\Psi_{\mathbf Q}|^2+\frac{b}{2}|\Psi_{\mathbf Q}|^4
+\frac{1}{2m^\ast}
\left(\mathbf Q^2-4\mathbf Q\cdot\mathbf A+4\mathbf A^2\right)
|\Psi_{\mathbf Q}|^2.
\label{3}
\end{align}

\noindent
In the absence of electromagnetic fields ($\mathbf A=\mathbf 0$), the equilibrium state is determined by
\begin{align}
\left.
\frac{\delta F_{\rm GL}[\Psi]}{\delta \Psi_{\mathbf Q}^{\ast}}
\right|_{\mathbf A=\mathbf 0}=0.
\label{4}
\end{align}

\noindent
Setting $\mathbf A=\mathbf 0$ in Eq.~(\ref{3}), we obtain
\begin{align}
\frac{F_{\rm GL}}{V}
=\left[\alpha(T)+\frac{\mathbf Q^2}{2m^\ast}\right]|\Psi_{\mathbf Q}|^2
+\frac{b}{2}|\Psi_{\mathbf Q}|^4.
\end{align}
Minimization with respect to $\Psi_{\mathbf Q}^{\ast}$ yields
\begin{align}
\alpha(T)+\frac{\mathbf Q^2}{2m^\ast}+b|\Psi_{\mathbf Q}|^2=0,
\end{align}
which leads to
\begin{align}
|\Psi_{\mathbf Q}|^2
=-\frac{1}{b}\left[\alpha(T)+\frac{\mathbf Q^2}{2m^\ast}\right].
\end{align}
Dividing by the $\mathbf Q=\mathbf 0$ value,
$|\Psi_{\mathbf Q=\mathbf 0}|^2=-\alpha(T)/b$, we obtain
\begin{align}
\frac{|\Psi_{\mathbf Q}|^2}{|\Psi_{\mathbf Q=\mathbf 0}|^2}
=1-\xi_0^2(T)\mathbf Q^2,
\label{5}
\end{align}
where the coherence length is defined as
\begin{align}
\xi_0^2(T)\equiv \frac{1}{2m^\ast|\alpha(T)|}.
\end{align}


The GL order parameter is proportional to an appropriate average of the microscopic superconducting gap.
We approximate
\begin{align}
\Psi_{\mathbf Q}\approx \langle \Delta^{(\mathbf Q)}\rangle
=\frac{1}{D(0)}\sum_{n\mathbf k}\Delta^{(\mathbf Q)}_{n\mathbf k},
\end{align}
where $D(0)$ is the density of states at the Fermi level and
$\Delta^{(\mathbf Q)}_{n\mathbf k}$ denotes the $\mathbf Q$-component of the gap in band $n$ and momentum $\mathbf k$.


The supercurrent density is defined as the functional derivative of the free energy with respect to $\mathbf A$:
\begin{align}
\mathbf J
\equiv -\frac{1}{V}
\left.
\frac{\delta F_{\rm GL}[\Psi]}{\delta \mathbf A}
\right|_{\mathbf A=\mathbf 0}.
\end{align}
From Eq.~(\ref{3}), the $\mathbf A$-dependent contribution gives
\begin{align}
\mathbf J
=\frac{2}{m^\ast}|\Psi_{\mathbf Q}|^2\mathbf Q.
\end{align}
Using Eq.~(\ref{5}), we obtain
\begin{align}
\mathbf J
=\frac{2}{m^\ast}|\Psi_{\mathbf Q=\mathbf 0}|^2
\left(1-\xi_0^2(T)\mathbf Q^2\right)\mathbf Q.
\end{align}
Defining the London penetration depth $\lambda_L(T)$ through
\begin{align}
\frac{2}{m^\ast}|\Psi_{\mathbf Q=\mathbf 0}|^2
\equiv \frac{1}{2\mu_0\lambda_L^2(T)},
\end{align}
we finally arrive at
\begin{align}
\mathbf J
=\frac{1}{2\mu_0\lambda_L^2(T)}
\left\{1-\xi_0^2(T)\mathbf Q^2\right\}\mathbf Q.
\label{eq_j_macro}
\end{align}
The current has a maximum for $\nabla_{\mathbf{Q}} \mathbf{J} = 0$ which yields $Q_{\mathrm{max}} = 1/(\sqrt{3}\xi_0 (T))$ and the depairing current 
\begin{align}
    J_{\mathrm{dp}}(T)\equiv|\mathbf{J}_{Q_{\mathrm{max}}}| = \frac{1}{3\sqrt{3}\mu_0\lambda_{\mathrm{L}}^2(T)\xi_0(T)}\;.
\end{align}

The penetration depth can equivalently be expressed from the linear response of the magnitude of the supercurrent density $j(Q)\equiv |\mathbf J|$ at $Q\to 0$:
\begin{align}
\lambda_L
=
\left(
2\mu_0\left.
\frac{\partial j(Q)}{\partial Q}
\right|_{Q=0}
\right)^{-1/2}.
\end{align}
We note that in the present unit system,
$\mu_0=4\pi\alpha^2$, where $\alpha$ is the fine-structure constant.

\subsection{Density functional theory for superconductors with finite-momentum Cooper pairs}

In density functional theory for superconductors (SCDFT), the singlet order parameter is defined as an extended density, which is the expectation value of the product of annihilation operators, $\hat{\psi}_\sigma(\mathbf{r})$, as~\cite{PhysRevLett.60.2430}:
\begin{align}
  \chi(\mathbf{r},\mathbf{r}')\equiv\left\langle\hat{\psi}_\uparrow(\mathbf{r})\hat{\psi}_\downarrow(\mathbf{r}')\right\rangle
  =\sum_{n\mathbf{k}}\{
    u_{n\mathbf{k}}(\mathbf{r})v_{n\mathbf{k}}^*(\mathbf{r}')f\left(-E_{nk}\right)
   -u_{n\mathbf{k}}(\mathbf{r}')v_{n\mathbf{k}}^*(\mathbf{r})f(E_{nk})\},
\label{eq_op}
\end{align}
where $u_{n\mathbf{k}}(\mathbf{r})$, $v_{n\mathbf{k}}(\mathbf{r})$, and $E_{n\mathbf{k}}$ are the eigenvectors and eigenvalues of the Kohn--Sham--Bogoliubov--de Gennes (KSBdG) equation:
\begin{align}
  \left(\begin{matrix}
    -\frac{\boldsymbol{\nabla}^2}{2}+V_\mathrm{KS}(\mathbf{r})-\mu & {\hat{\Delta}}_\mathrm{KS} \\
    {\hat{\Delta}}_\mathrm{KS}^\dag & \frac{\boldsymbol{\nabla}^2}{2}-V_\mathrm{KS}(\mathbf{r})+\mu
  \end{matrix}\right)
  \left(\begin{matrix}
    u_{n\mathbf{k}}(\mathbf{r}) \\
    v_{n\mathbf{k}}(\mathbf{r})
  \end{matrix}\right)
  =E_{n\mathbf{k}}
  \left(\begin{matrix}
    u_{n\mathbf{k}}(\mathbf{r}) \\
    v_{n\mathbf{k}}(\mathbf{r})
  \end{matrix}\right).
\label{eq_ksbdg}
\end{align}
The effective diagonal and off-diagonal potentials, $V_\mathrm{KS}(\mathbf{r})$ and ${\hat{\Delta}}_\mathrm{KS}$, of the auxiliary non-interacting system are computed by taking the functional derivative of the exchange-correlation grand potential functional $\Omega_\mathrm{xc}[\rho, \chi]$ with respect to the normal density $\rho(\mathbf{r})$ and the singlet order parameter $\chi(\mathbf{r},\mathbf{r}')$ as:
\begin{align}
V_\mathrm{KS}(\mathbf{r}) = \frac{\delta \Omega_\mathrm{xc}}{\delta \rho(\mathbf{r})},
\qquad
{\hat{\Delta}}_\mathrm{KS}\psi(\mathbf{r}) \equiv \int d^3 r' \Delta_\mathrm{KS}(\mathbf{r},\mathbf{r}') \psi(\mathbf{r}'),
\qquad \Delta_\mathrm{KS}(\mathbf{r},\mathbf{r}') = -\frac{\delta \Omega_\mathrm{xc}}{\delta \chi^*(\mathbf{r},\mathbf{r}')}
\end{align}
By combining SCDFT and current density functional theory~\cite{PhysRevLett.59.2360,PhysRevB.37.10685},
the current density in the superconducting state is computed as~\cite{doi:10.7566/JPSJ.86.104705}:
\begin{align}
  \mathbf{j}(\mathbf{r}) &=
  \frac{i}{2}\sum_{\sigma}\left\langle\hat{\psi}_\sigma^\dag(\mathbf{r})\boldsymbol{\nabla}\hat{\psi}_\sigma(\mathbf{r})-\hat{\psi}_\sigma(\mathbf{r})\boldsymbol{\nabla}\hat{\psi}_\sigma^\dag(\mathbf{r})\right\rangle
  \\
  &= i\sum_{\mathbf{k}}\left[u_{n\mathbf{k}}^*(\mathbf{r})\boldsymbol{\nabla} u_{n\mathbf{k}}(\mathbf{r})f(E_{n\mathbf{k}})+v_{n\mathbf{k}}(\mathbf{r})\boldsymbol{\nabla} v_{n\mathbf{k}}^*(\mathbf{r})f(E_{n\mathbf{k}})\right]+c.c.
\label{eq_cdensity}
\end{align}

\begin{figure}[tbp!]
\includegraphics[width=1\textwidth]{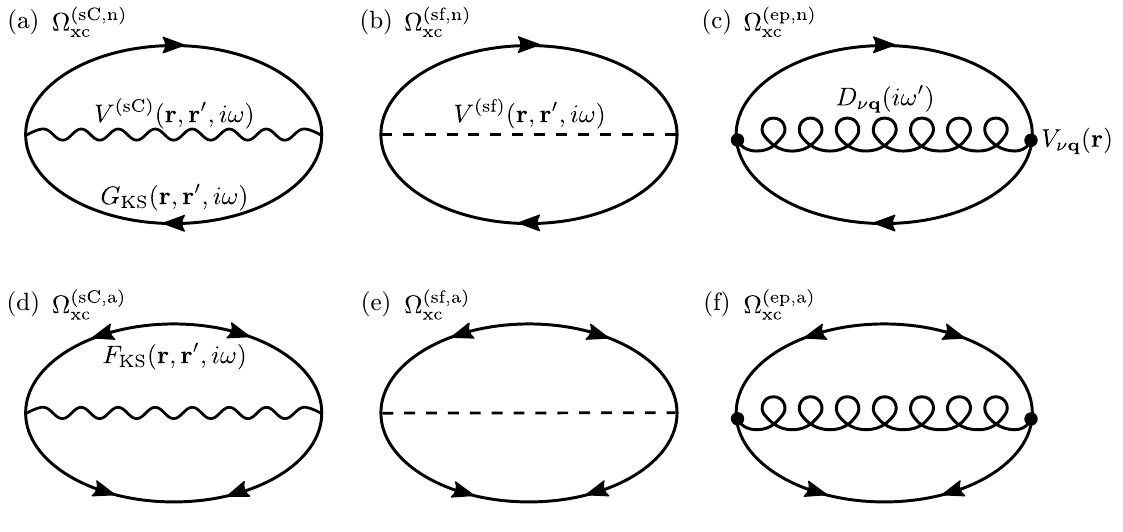}
\caption{\label{fig:diagram} 
Feynman diagrams of the normal (a-c) and anomalous (d-f) part of the exchange-correlation grandpotentaial.
``sC'', ``sf'', and ``ep'' denote the screened Coulomb (a, d), spin-fluctuation (b, e), and electron-phonon (c, f) contributions, respectively.
}
\end{figure}

First, we explain the ordinary lattice periodic case.
In that condition, the singlet order parameter has the lattice translational invariance as:
\begin{align}
  \chi\left(\mathbf{r}+\mathbf{R},\mathbf{r}'+\mathbf{R}\right)=\chi(\mathbf{r},\mathbf{r}'),
\end{align}
where $\mathbf{R}$ denotes an arbitrary lattice vector.
If the energy scale of the off-diagonal component of KSBdG eq. [Eq.~(\ref{eq_ksbdg})] is much smaller than that scale of the diagonal component, we ignore the hybridization of bands caused by ${\hat{\Delta}}_\mathrm{KS}$ and apply the decoupling approximation~\cite{PhysRevB.72.024546} as:
\begin{align}
\left\{
\begin{aligned}
  u_{n\mathbf{k}}(\mathbf{r}) &\approx u_{n\mathbf{k}}\varphi_{n\mathbf{k}}(\mathbf{r}),
  \\
  v_{n\mathbf{k}}(\mathbf{r}) &\approx v_{n\mathbf{k}}\varphi_{n\mathbf{k}}(\mathbf{r}),
\end{aligned}
\right.
\label{eq_decouple}
\end{align}
where $\varphi_{n\mathbf{k}}(\mathbf{r})$ is the eigenfunction of the normal-state Kohn--Sham eq.,
\begin{align}
\left(-\frac{\boldsymbol{\nabla}^2}{2}+V_\mathrm{KS}(\mathbf{r})\right)\varphi_{n\mathbf{k}}(\mathbf{r}) 
= \xi_{n\mathbf{k}}\varphi_{n\mathbf{k}}(\mathbf{r}) .
\end{align}
This leads the KSBdG eq. to a $2\times2$ secular equation
\begin{align}
  \left(\begin{matrix}
    \xi_{n\mathbf{k}} & \Delta_{n\mathbf{k}} \\
    \Delta_{n\mathbf{k}}^* & -\xi_{n\mathbf{k}}
  \end{matrix}\right)
  \left(\begin{matrix}
    u_{n\mathbf{k}} \\
    v_{n\mathbf{k}}
  \end{matrix}\right)
  =E_{n\mathbf{k}}\left(\begin{matrix}
    u_{n\mathbf{k}} \\
    v_{n\mathbf{k}} \\
  \end{matrix}\right),
\end{align}
where $E_{n\mathbf{k}}=\sqrt{\xi_{n\mathbf{k}}^2+|\Delta_{n\mathbf{k}}|^2}$, and $\Delta_{n\mathbf{k}}\equiv\Delta_{nn}(\mathbf{k})$ is the band-diagonal part of the following matrix:
\begin{align}
\Delta_{nn'}(\mathbf{k}) &\equiv \iint d^3r d^3 r' 
  \varphi^*_{n\mathbf{k}}(\mathbf{r}) \Delta_\mathrm{KS}(\mathbf{r},\mathbf{r}') \varphi_{n'\mathbf{k}}(\mathbf{r}').
\end{align}
In deriving Eq.~(\ref{eq_decouple}), we employ the standard decoupling approximation of SCDFT, in which the KSBdG amplitudes are expanded using the corresponding normal-state Kohn--Sham orbital and the anomalous potential is retained only within the same band.
Equivalently, off-diagonal pairing matrix elements $\Delta_{nn'}(\mathbf{k})$ with $n\neq n'$ are neglected.
The validity of this approximation should not be formulated as the requirement that the superconducting gap be smaller than the separation between every pair of neighboring Kohn--Sham bands throughout the Brillouin zone. Local accidental or symmetry-enforced degeneracies do not necessarily invalidate the approximation.
The relevant assumption is instead that superconductivity is dominated by intraband pairing and that interband anomalous matrix elements remain small or are important only in a restricted region of momentum space.
The decoupling approximation may become insufficient when nearly degenerate bands are strongly coupled by the pairing potential over an extended momentum-space region. Possible examples include low-symmetry systems with strong spin-orbit coupling, some non-centrosymmetric superconductors, and multiorbital systems with substantial interband pairing.
In such cases, the full multiband KSBdG equation should be solved without restricting the anomalous potential to its band-diagonal components.
For the conventional phonon-mediated superconductors considered in this work, the decoupling approximation is well established.
It has been used in previous SCDFT calculations~\cite{PhysRevLett.125.057001} to reproduce transition temperatures, superconducting gaps of weak- and strong-coupling elemental superconductors and related conventional compounds.
We therefore expect the neglected interband pairing contributions to have only a minor effect on the quantities evaluated here.
Then, the singlet order parameter in Eq.~(\ref{eq_op}) becomes
\begin{align}
\chi(\mathbf{r},\mathbf{r}') &= \sum_{n\mathbf{k}} \varphi_{n\mathbf{k}}(\mathbf{r}) \varphi^*_{n\mathbf{k}}(\mathbf{r}') \chi_{n\mathbf{k}},
\\
\chi_{n\mathbf{k}} &\equiv \iint d^3r d^3 r' \varphi^*_{n\mathbf{k}}(\mathbf{r}) \chi(\mathbf{r},\mathbf{r}') \varphi_{n\mathbf{k}}(\mathbf{r}')
=
\frac{\Delta_{n\mathbf{k}}}{2 E_{n\mathbf{k}}} \tanh\left(\frac{E_{n\mathbf{k}}}{2 T}\right).
\end{align}
$\Delta_{n\mathbf{k}}$ satisfies the following Kohn--Sham gap eq.:
\begin{align}
  \Delta_{n\mathbf{k}} &\equiv \iint d^3r d^3 r' 
  \varphi^*_{n\mathbf{k}}(\mathbf{r}) \Delta_\mathrm{KS}(\mathbf{r},\mathbf{r}') \varphi_{n\mathbf{k}}(\mathbf{r}')
  =
  -\frac{\delta \Omega_\mathrm{xc}\left[\{\chi_{n\mathbf{k}}(\Delta_{n\mathbf{k}})\},\{\chi^*_{n\mathbf{k}}(\Delta_{n\mathbf{k}})\}\right]}{\delta \chi^*_{n\mathbf{k}}}
 \label{ks_gap_eq0}
\nonumber \\
&=-Z_{n\mathbf{k}} \Delta_{n\mathbf{k}}
  -\frac{1}{2} \sum_{n'\mathbf{k}'}
  K_{n\mathbf{k}n'\mathbf{k}'} \frac{\Delta_{n'\mathbf{k}'}}{E_{n'\mathbf{k}'}}
  \tanh\left(\frac{E_{n'\mathbf{k}'}}{2T}\right),
\end{align}
where we employed partially linearized form of the gap equation~\cite{PhysRevB.72.024545}, and approximate $\Omega_\mathrm{xc}$ by the Kohn--Sham perturbation theory with the lowest-order diagrams shown in Fig.~\ref{fig:diagram} as
\begin{align}
\Omega_\mathrm{xc} \approx \Omega^\textrm{(sC,n)}_\mathrm{xc} + \Omega^\textrm{(sf,n)}_\mathrm{xc} + \Omega^\textrm{(ep,n)}_\mathrm{xc}
+\Omega^\textrm{(sC,a)}_\mathrm{xc} + \Omega^\textrm{(sf,a)}_\mathrm{xc} + \Omega^\textrm{(ep,a)}_\mathrm{xc},
\end{align}
and these grandpotentials are constructed the normal-, anomalous, and phonon Kohn--Sham (unperturbed) Green's functions as
\begin{align}
G^\textrm{(KS)}_{\sigma\sigma'}(\textbf{r},\textbf{r}',i\omega) &= -\int_0^{1/T} d\tau e^{i \omega \tau}
 \langle \hat{\mathcal{T}} \hat{\psi}_\sigma(\textbf{r},\tau)\hat{\psi}^\dagger_{\sigma'}(\textbf{r}',0)\rangle_\textrm{KS} ,
\\
F^\textrm{(KS)}_{\sigma\sigma'}(\textbf{r},\textbf{r}',i\omega) &= -\int_0^{1/T} d\tau e^{i \omega \tau}
 \langle \hat{\mathcal{T}} \hat{\psi}_\sigma(\textbf{r},\tau)\hat{\psi}_{\sigma'}(\textbf{r}',0)\rangle_\textrm{KS} ,
\\
D_{\nu\textbf{q}}(i\omega) &= \frac{2\omega_{\nu\textbf{q}}}{\omega^2+\omega_{\nu\textbf{q}}^2},
\end{align}
respectively.
The renormalization factor $Z_{n\textrm{k}}$ and the kernel $K_{n\mathbf{k}n'\mathbf{k}'}$ are computed from the normal- and anomalous- term of the grandpotentials as:
\begin{align}
 Z_{n\mathbf{k}} \equiv Z^{\mathrm{(sC)}}_{n\mathbf{k}} + Z^{\mathrm{(sf)}}_{n\mathbf{k}} + Z^\mathrm{(ep)}_{n\mathbf{k}}
\equiv
\left. \left(
\frac{1}{\Delta_{n\textbf{k}}}\frac{\delta(\Omega^\textrm{(sC,n)}_\mathrm{xc} + \Omega^\textrm{(sf,n)}_\mathrm{xc} + \Omega^\textrm{(ep,n)}_\mathrm{xc})}{\delta \chi^*_{n\textbf{k}}}
\right)\right|_{\{\Delta_{n\textbf{k}}\}=0},
\label{eq_z_derivative}
\\
K_{n\mathbf{k}n'\mathbf{k}'}\equiv 
K^\mathrm{(sC)}_{n\mathbf{k}n'\mathbf{k}'} + K^\mathrm{(sf)}_{n\mathbf{k}n'\mathbf{k}'} + K^\mathrm{(ep)}_{n\mathbf{k}n'\mathbf{k}'}
\equiv
\left. 
\frac{\delta^2
\left(\Omega^\textrm{(sC,a)}_\mathrm{xc} + \Omega^\textrm{(sf,a)}_\mathrm{xc} + \Omega^\textrm{(ep,a)}_\mathrm{xc}\right)}
{\delta \chi^*_{n\textbf{k}}\delta \chi_{n'\textbf{k}'}}
\right|_{\{\Delta_{n\textbf{k}}\}=0}.
\label{eq_k_derivative}
\end{align}
With this condition, the screened-Coulomb (sC) and spin-fluctuation (sf) part becomes
\begin{align}
Z^\mathrm{(sC)/(sf)}_{n\mathbf{k}} &=\mp\sum_{n'\mathbf{k}'}\frac{1}{\pi}
\int_0^\infty d\omega \frac{(|\xi_{n\mathbf{k}}|+|\xi_{n'\mathbf{k}'}|)^2-\omega^2}
{\{(|\xi_{n\mathbf{k}}|+|\xi_{n'\mathbf{k}'}|)^2+\omega^2\}^2}
V^\mathrm{(sC)/(sf)}_{n\mathbf{k}n'\mathbf{k}'}(i\omega),
\\
K_{n\mathbf{k}n'\mathbf{k}'}^\mathrm{(sC)/(sf)}
&=\frac{2}{\pi} \int_0^\infty d\omega  
\frac{|\xi_{n\mathbf{k}}|+|\xi_{n'\mathbf{k}'}|}{(|\xi_{n\mathbf{k}}|+|\xi_{n'\mathbf{k}'}|)^2+\omega^2}
 V^\mathrm{(sC)/(sf)}_{n\mathbf{k}n'\mathbf{k}'}(i\omega),
\end{align}
where $-(+)$ is used for the sC (sf) case, and the exchange integral with effective potential
\begin{align}
V^\mathrm{(sC)/(sf)}_{n\mathbf{k}n'\mathbf{k}'}(i\omega)
&=\iint d^3r d^3r' \rho_{n\mathbf{k}n'\mathbf{k}'}(\mathbf{r}) \rho^*_{n\mathbf{k}n'\mathbf{k}'}(\mathbf{r}')
V^\mathrm{(sC)/(sf)}(\mathbf{r},\mathbf{r}',i\omega),
\\
\rho_{n\mathbf{k}n'\mathbf{k}'} (\mathbf{r})&=\varphi_{n\mathbf{k}}^* (\mathbf{r}) 
\varphi_{n' \mathbf{k}'} (\mathbf{r})+\sum_{I l l'} Q_{I l l'} (\mathbf{r})\langle\varphi_{n\mathbf{k}}|\beta_{I l} \rangle
  \langle\beta_{I l'}|\varphi_{n' \mathbf{k}'} \rangle
\end{align}
is evaluated by including the augmentation charge~\cite{PhysRevB.73.045112} $Q_{I l l'} (\mathbf{r}) = \varphi_{I l}^\mathrm{ae *} (\mathbf{r})\varphi_{I l'}^\mathrm{ae} (\mathbf{r})
-\varphi_{I l}^\mathrm{ps *} (\mathbf{r})\varphi_{I l'}^\mathrm{ps} (\mathbf{r})$
when we employed the ultrasoft pseudopotential~\cite{PhysRevB.41.7892} with $\varphi_{I l}^\mathrm{ae}$, $\varphi_{I l}^\mathrm{ps}$ and $\beta_{I l}$ as the $l$-th all-electron, pseudo, and its dual orbitals of atom $I$, respectively.
The screened Coulomb potential is computed with the random-phase approximation as~\cite{PhysRevLett.111.057006,PhysRevB.102.214508}:
\begin{align}
{\hat V}^\mathrm{(sC)}(i\omega)=\{{\hat 1}-{\hat V}_\mathrm{bare} {\hat \Pi}_0(i\omega)\}^{-1} {\hat V}_\mathrm{bare},
\end{align}
where we wrote an $\textbf{r}$-space matrix as $V(\textbf{r}, \textbf{r}')\equiv\hat{V}$.
$V_\mathrm{bare}(\textbf{r}, \textbf{r}')=1/|\textbf{r}-\textbf{r}'|$ is the bare Coulomb interaction, and
\begin{align}
{\hat \Pi}_0 (\mathbf{r},\mathbf{r}',i\omega)=\sum_{n\mathbf{k}n'\mathbf{k}'}
\frac{\theta(-\xi_{n\mathbf{k}})-\theta(-\xi_{n'\mathbf{k}'})}{\xi_{n\mathbf{k}}-\xi_{n'\mathbf{k}'}+i\omega}
\rho_{n\mathbf{k}n'\mathbf{k}'} (\mathbf{r}) \rho^*_{n\mathbf{k}n'\mathbf{k}'} (\mathbf{r}')
\end{align}
is the susceptibility of the Kohn--Sham non-interacting system.
The spin-fluctuation mediated interaction is computed with the magnetic exchange-correlation kernel ${\hat I}_\mathrm{xc}$ as~\cite{PhysRevB.90.214504}
\begin{align}
{\hat V}^\mathrm{(sf)}(i\omega)&=-3{\hat I}_\mathrm{xc} {\hat \chi}_\mathrm{s}(i\omega) {\hat I}_\mathrm{xc}
\\
{\hat \chi}_\mathrm{s}(i\omega)&=\{{\hat 1}-{\hat I}_\mathrm{xc} {\hat \Pi}_0(i\omega)\}^{-1} {\hat \Pi}_0(i\omega)
\\
{\hat I}_\mathrm{xc}&\equiv\frac{\delta^2 E_\mathrm{xc}}{\delta m(\mathbf{r})\delta m(\mathbf{r}')}.
\end{align}
In this study, we employed the adiabatic local-density approximation (ALDA) for the exchange-correlation kernel.
This corresponds to the uniform electron-gas limit of the direct $T$-matrix treatment used in the fluctuation-exchange approximation for the extended Hubbard model.
The electron-phonon is computed by including the Eliashberg-combining correction~\cite{PhysRevLett.125.057001} as:
\begin{align}
Z^\mathrm{(ep)}_{n\mathbf{k}} &= - \frac{1}{\tanh(\xi_{n\mathbf{k}}/(2T))} \sum_{n'\mathbf{k}'\nu} |g^\nu_{n\mathbf{k}n'\mathbf{k}'}|^2
\frac{d}{d\xi_{n\mathbf{k}}} \{
  I_T(\xi_{n\mathbf{k}},\xi_{n'\mathbf{k}'},\omega_{\nu \mathbf{k}'-\mathbf{k}})
+ I_T(\xi_{n\mathbf{k}},-\xi_{n'\mathbf{k}'},\omega_{\nu \mathbf{k}'-\mathbf{k}})\},
\\
K^\mathrm{(ep)}_{n\mathbf{k}n'\mathbf{k}'} &= \frac{\gamma_1 \gamma_2}{\tanh(\xi_{n\mathbf{k}}/(2T)) \tanh(\xi_{n'\mathbf{k}'}/(2T))} \omega_{\nu \mathbf{k}'-\mathbf{k}} |g^\nu_{n\mathbf{k}n'\mathbf{k}'}|^2
\sum_{s_i = \pm} s_1 s_2 s_3 J(\xi_{n\mathbf{k}}, s_1 \xi_{n'\mathbf{k}'}, s_2 \omega_{\nu \mathbf{k}'-\mathbf{k}}, s_3 \gamma_2 \omega_{\nu \mathbf{k}'-\mathbf{k}}),
\\
I_T(\xi,\xi',\omega) &= f_T(\xi)f_T(\xi')n_T(\omega) \left[ \frac{e^{\xi/T}-e^{(\xi'+\omega)/T}}{\xi-\xi'-\omega}
-\frac{e^{\xi'/T}-e^{(\xi+\omega)/T}}{\xi-\xi'+\omega} \right],
\\
J_T(\xi,E,\omega,\gamma) &= [f_T(\xi)+n_T(\omega)] \left[
\frac{f_T(\xi-\omega)}{(\xi-\omega-E)(\xi-\omega-\gamma)}
+\frac{f_T(E)}{(\gamma-E)(\xi-\omega-E)}
-\frac{f_T(\gamma)}{(\gamma-E)(\xi-\omega-\gamma)}
\right],
\end{align}
where $f_T(\xi)$ and $n_T(\omega)$ are the Fermi--Dirac and Bose--Einstein distribution function respectively.
Parameter $\gamma_1 = 1.33$ and $\gamma_2 = 3.8$ are tuned so that the result coincides with the Eliashberg theory when only the electron-phonon part is considered.
The electron-phonon vertex $g^\nu_{n\mathbf{k}n'\mathbf{k}'}$ is computed with the screened (Kohn--Sham) deformation potential $V_{\nu\textbf{q}}$ as:
\begin{align}
g^\nu_{n\mathbf{k}n'\mathbf{k}'} = 
 \int d^3 r
    \varphi_{n\textbf{k}}^*(\textbf{r}) \varphi_{n'\textbf{k}'}(\textbf{r})
V_{\nu\textbf{k}'-\textbf{k}}(\textbf{r}).
\end{align}

To treat superconductors with a finite-momentum order parameter, we impose twisted boundary conditions as follows~\cite{PhysRev.135.A550}~(see Supplemental Material of Ref.~\cite{Witt2024}):
\begin{align}
  \chi(\mathbf{r}+\mathbf{R},\mathbf{r}'+\mathbf{R})=e^{i\mathbf{Q}\cdot\mathbf{R}}\chi(\mathbf{r},\mathbf{r}').
\end{align}
Similarly to the treatment of the spin-spiral state~\cite{sandratskii1998noncollinear,PhysRevB.96.195133}, this corresponds to a generalized Bloch theorem~\cite{Witt2024} and we can use the decoupling approximation to obtain:
\begin{align}
  v_{n\mathbf{k}}(\mathbf{r})\approx v_{n\mathbf{k}}\varphi_{n\mathbf{k}-\mathbf{Q}/2}(\mathbf{r}),
  \\
  u_{n\mathbf{k}}(\mathbf{r})\approx u_{n\mathbf{k}}\varphi_{n\mathbf{k}+\mathbf{Q}/2}(\mathbf{r}).
\end{align}
This modifies the $2\times2$ secular equation to:
\begin{align}
  \left(\begin{matrix}
\xi_{n\mathbf{k}+\mathbf{Q}/2}&\Delta^{(\mathbf{Q})}_{n\mathbf{k}}\\
\Delta^{(\mathbf{Q})*}_{n\mathbf{k}}&-\xi_{n\mathbf{k}-\mathbf{Q}/2}\\
\end{matrix}\right)\left(
\begin{matrix}u_{n\mathbf{k}}\\v_{n\mathbf{k}}\\
\end{matrix}\right)=E_{n\mathbf{k}}\left(\begin{matrix}u_{n\mathbf{k}}\\v_{n\mathbf{k}}\\\end{matrix}\right),
\\
\Delta^{(\mathbf{Q})}_{n\mathbf{k}} \equiv \int d^3r \int d^3r' \varphi^*_{n\mathbf{k}+\mathbf{Q}/2}(\mathbf{r})
\Delta(\mathbf{r},\mathbf{r}') \varphi^*_{n\mathbf{k}-\mathbf{Q}/2}(\mathbf{r}').
\end{align}
Then, the singlet order parameter in Eq.~(\ref{eq_op}) becomes
\begin{align}
\chi(\mathbf{r},\mathbf{r}') &= \sum_{n\mathbf{k}} \varphi_{n\mathbf{k}+\textbf{Q}/2}(\mathbf{r})
\varphi^*_{n\mathbf{k}-\textbf{Q}/2}(\mathbf{r}') \chi^{(\mathbf{Q})}_{n\mathbf{k}},
\\
\chi^{(\mathbf{Q})}_{n\mathbf{k}} &\equiv \iint d^3r d^3 r' \varphi^*_{n\mathbf{k}+\textbf{Q}/2}(\mathbf{r})
\chi(\mathbf{r},\mathbf{r}') \varphi_{n\mathbf{k}-\textbf{Q}/2}(\mathbf{r}')
=
\frac{\Delta^{(\mathbf{Q})}_{n\mathbf{k}}}{4 E_{n\mathbf{k}}^{(\mathbf{Q})}} 
  \left\{
    \tanh\left(\frac{E_{n\mathbf{k}}^{(\mathbf{Q})}+d_{n\mathbf{k}}^{(\mathbf{Q})}}{2T}\right)
   +\tanh\left(\frac{E_{n\mathbf{k}}^{(\mathbf{Q})}-d_{n\mathbf{k}}^{(\mathbf{Q})}}{2T}\right)
  \right\},
\end{align}
where 
\begin{align}
  E_{n\mathbf{k}}^{(\mathbf{Q})} \equiv\sqrt{{\bar{\xi}}_{n\mathbf{k}}^2+\left|\Delta_{n\mathbf{k}}^{(\mathbf{Q})}\right|^2},
  \qquad
  {\bar{\xi}}_{n\mathbf{k}}^{(\mathbf{Q})} \equiv\frac{\xi_{n\mathbf{k}+\mathbf{Q}/2}+\xi_{n\mathbf{k}-\mathbf{Q}/2}}{2},
  \qquad
  d_{n\mathbf{k}}^{(\mathbf{Q})} \equiv\frac{\xi_{n\mathbf{k}+\mathbf{Q}/2}-\xi_{n\mathbf{k}-\mathbf{Q}/2}}{2}.
\end{align}
In this case, the Kohn--Sham gap equation~(\ref{ks_gap_eq0}) becomes
\begin{align}
  \Delta_{n\mathbf{k}}^{(\mathbf{Q})}= -Z^{(\mathbf{Q})}_{n\mathbf{k}} \Delta_{n\mathbf{k}}^{(\mathbf{Q})}
  -\frac{1}{4}\sum_{n'\mathbf{k}'} K^{(\mathbf{Q})}_{n\mathbf{k}n'\mathbf{k}'}
  \frac{\Delta_{n'\mathbf{k}'}^{(\mathbf{Q})}}{E_{n'\mathbf{k}'}^{(\mathbf{Q})}}
  \left\{
    \tanh\left(\frac{E_{n'\mathbf{k}'}^{(\mathbf{Q})}+d_{n'\mathbf{k}'}^{(\mathbf{Q})}}{2T}\right)
   +\tanh\left(\frac{E_{n'\mathbf{k}'}^{(\mathbf{Q})}-d_{n'\mathbf{k}'}^{(\mathbf{Q})}}{2T}\right)
  \right\},
\label{eq_gap_q}
\end{align}
Let us estimate The $\textbf{Q}$-dependence of renormalization factor $Z^{(\mathbf{Q})}_{n\mathbf{k}}$ and kernel $K^{(\mathbf{Q})}_{n\mathbf{k}n'\mathbf{k}'}$, that are obtained by the functional derivative in Eqs~(\ref{eq_z_derivative}) and (\ref{eq_k_derivative}). 
As a typical case, we consider the static part in the diagram Fig.~\ref{fig:diagram}~(d),
\begin{align}
\Omega_\textrm{xc}^\textrm{sC,a,static} = \int d^3r \int d^3r'
\chi(\textbf{r},\textbf{r}') V^\textrm{(sC)}(\textbf{r},\textbf{r},0) \chi^*(\textbf{r},\textbf{r}'),
\end{align}
and the corresponding kernel,
\begin{align}
K_{n\textbf{k}n'\textbf{k}'}^{(\textbf{Q},\textrm{sC,static})} \equiv
\frac{\delta^2 \Omega_\textrm{xc}^\textrm{sC,a,static}}{\delta \chi^{(\textbf{Q})*}_{n\textbf{k}}\chi^{(\textbf{Q})}_{n'\textbf{k}'}}
= \int d^3r \int d^3r' \varphi^*_{n\mathbf{k}+\textbf{Q}/2}(\mathbf{r}) \varphi_{n\mathbf{k}-\textbf{Q}/2}(\mathbf{r}')
 V^\textrm{(sC)}(\textbf{r},\textbf{r}',0)
\varphi_{n'\mathbf{k}'+\textbf{Q}/2}(\mathbf{r}) \varphi^*_{n'\mathbf{k}'-\textbf{Q}/2}(\mathbf{r}').
\end{align}
If we approximate $\varphi_{n\mathbf{k}}(\mathbf{r})$ with an uniform electron gas $\exp(i\textbf{k}\cdot\textbf{r})/\sqrt{V}$,
\begin{align}
K_{n\textbf{k}n'\textbf{k}'}^{(\textbf{Q},\textrm{sC,static})} \approx
\frac{1}{V^2}\int d^3r \int d^3r' V^\textrm{(sC)}(\textbf{r},\textbf{r}',0) e^{i (\textbf{k}'-\textbf{k})\cdot\textbf{r}}
e^{i (\textbf{k}-\textbf{k}')\cdot\textbf{r}'}
\end{align}
has no $\textbf{Q}$-dependence.
As an opposite limit, we assume the $\varphi_{n\mathbf{k}}(\mathbf{r})$ as a superposition of the localized orbital $N_\textbf{R}^{-1/2}\sum_\textbf{R} \exp(i \textbf{k}\cdot\textbf{R}) \varphi_n(\textbf{r}-\textbf{R})$,
\begin{align}
K_{n\textbf{k}n'\textbf{k}'}^{(\textbf{Q},\textrm{sC,static})} = 
\sum_{\textbf{R}_1 \textbf{R}_2 \textbf{R}_3 \textbf{R}_4}
\frac{1}{N_\textbf{R}^2}\int d^3r \int d^3r' V^\textrm{(sC)}(\textbf{r},\textbf{r}',0)
\varphi^*_n(\textbf{r}-\textbf{R}_1)\varphi_{n'}(\textbf{r}-\textbf{R}_2)\varphi^*_{n'}(\textbf{r}'-\textbf{R}_3)\varphi_n(\textbf{r}'-\textbf{R}_4)
\nonumber \\
\times \exp\left[i \left\{-\left(\textbf{k}+\frac{\textbf{Q}}{2}\right)\cdot\textbf{R}_1 
+ \left(\textbf{k}'+\frac{\textbf{Q}}{2}\right)\cdot\textbf{R}_2
-\left(\textbf{k}'-\frac{\textbf{Q}}{2}\right)\cdot\textbf{R}_3
+\left(\textbf{k}-\frac{\textbf{Q}}{2}\right)\cdot\textbf{R}_4\right\}\right].
\end{align}
If the orbitals $\varphi_n$ are localized and overlap only at $\textbf{R}_1=\textbf{R}_2$ and $\textbf{R}_3=\textbf{R}_4$, $K_{n\textbf{k}n'\textbf{k}'}^{(\textbf{Q},\textrm{sC,static})}$ becomes a two-center integral as
\begin{align}
K_{n\textbf{k}n'\textbf{k}'}^{(\textbf{Q},\textrm{sC,static})} \approx
\sum_{\textbf{R} \textbf{R}'}
\frac{1}{N_\textbf{R}^2}\int d^3r \int d^3r' V^\textrm{(sC)}(\textbf{r},\textbf{r}',0)
\varphi^*_n(\textbf{r}-\textbf{R})\varphi_{n'}(\textbf{r}-\textbf{R})\varphi^*_{n'}(\textbf{r}'-\textbf{R}')\varphi_n(\textbf{r}'-\textbf{R}')
e^{i (\textbf{R}' - \textbf{R})\cdot(\textbf{k}-\textbf{k}')},
\end{align}
and is also $\textbf{Q}$-independent.
These two limiting cases show that the $\mathbf{Q}$ dependence vanishes both in the uniform-electron-gas limit and in the strictly localized-orbital limit.
For realistic Bloch states, a residual $\mathbf{Q}$ dependence may arise from finite intersite overlap and from the variation of the interaction matrix elements.
In the present calculations, however, the relevant momenta satisfy $|\mathbf Q|a\ll1$.
We therefore employ the long-wavelength approximation
\begin{align}
  K_{n\textbf{k}n'\textbf{k}'}^{(\textbf{Q})}\simeq K_{n\textbf{k}n'\textbf{k}'}^{(\textbf{0})}
=K_{n\textbf{k}n'\textbf{k}'},
\qquad
Z_{n\textbf{k}}^{(\textbf{Q})}\simeq Z_{n\textbf{k}}^{(\textbf{0})}=Z_{n\textbf{k}}.
\end{align}
The two limiting cases above provide physical support for this approximation, although a fully \(Q\)-dependent evaluation of the kernels remains a subject for future work.
With this approximation, Eq.~(\ref{eq_gap_q}) becomes
\begin{align}
  \Delta_{n\mathbf{k}}^{(\mathbf{Q})}=-Z_{n\mathbf{k}} \Delta_{n\mathbf{k}}^{(\mathbf{Q})}
  -\frac{1}{4}\sum_{n'\mathbf{k}'} K_{n\mathbf{k}n'\mathbf{k}'}
  \frac{\Delta_{n'\mathbf{k}'}^{(\mathbf{Q})}}{E_{n'\mathbf{k}'}^{(\mathbf{Q})}}
  \left\{
    \tanh\left(\frac{E_{n'\mathbf{k}'}^{(\mathbf{Q})}+d_{n'\mathbf{k}'}^{(\mathbf{Q})}}{2T}\right)
   +\tanh\left(\frac{E_{n'\mathbf{k}'}^{(\mathbf{Q})}-d_{n'\mathbf{k}'}^{(\mathbf{Q})}}{2T}\right)
  \right\}.
\label{eq_gap_q_kz}
\end{align}
The Kohn--Sham energy at a $\mathbf{k}$ point slightly displaced from the grid point is computed with the Taylor expansion as:
\begin{align}
  \xi_{n\mathbf{k}+\mathbf{Q}/2} \approx \xi_{n\mathbf{k}}
  +\sum_{\alpha=x,y,z}{\frac{\partial\xi_{n\mathbf{k}}}{{\partial k}_\alpha}\frac{Q_\alpha}{2}}+\frac{1}{2}
  \sum_{\alpha,\alpha'=x,y,z}{\frac{\partial^2\xi_{n\mathbf{k}}}{{\partial k}_\alpha{\partial k}_{\alpha'}}\frac{Q_\alpha}{2}\frac{Q_{\alpha'}}{2}},
\end{align}
where the first and second derivatives are computed using finite differences.

Because the integrand of the gap equation [Eq.~(\ref{eq_gap_q_kz})] changes rapidly in the vicinity of the Fermi surface, we use an auxiliary energy-grid scheme~\cite{PhysRevB.95.054506} to stabilize the $\mathbf{k}$-point integration as follows: 
\begin{align}
  \Delta_{n\mathbf{k}}^{(\mathbf{Q})}(\xi)=
  -\frac{1}{2}\int{d\xi'} \sum_{n'\mathbf{k}'} \delta(\xi'-{\bar{\xi}}_{n'\mathbf{k}'}^{(\mathbf{Q})})
  &\frac{K_{n\mathbf{k}n'\mathbf{k}'}(\xi,\xi')}{1+Z_{n\mathbf{k}}(\xi)}
  \frac{\Delta_{n'\mathbf{k}'}^{(\mathbf{Q})}(\xi')}{E_{n'\mathbf{k}'}^{(\mathbf{Q})}(\xi')}
  \tanh{\left(\frac{E_{n'\mathbf{k}'}^{(\mathbf{Q})}(\xi')}{2T}\right)}
  \nonumber \\
  &\times
  \frac{1-\tanh^2{\left(\frac{d_{n'\mathbf{k}'}^{(\mathbf{Q})}(\xi')}{2T}\right)}}
    {1-\tanh^2{\left(\frac{E_{n'\mathbf{k}'}^{(\mathbf{Q})}(\xi')}{2T}\right)}\tanh^2{\left(\frac{d_{n'\mathbf{k}'}^{(\mathbf{Q})}(\xi')}{2T}\right)}},
\end{align}
where $E_{n\mathbf{k}}^{(\mathbf{Q})}(\xi)\equiv\sqrt{\xi^2 + |\Delta_{n\mathbf{k}}^{(\mathbf{Q})}(\xi)|}$, and
\begin{align}
  d_{n\mathbf{k}}^{(\mathbf{Q})}(\xi)\equiv
  \frac{\left\langle
    \left(\xi_{n\mathbf{k}+\mathbf{Q}/2}-\xi_{n\mathbf{k}-\mathbf{Q}/2}\right)\delta(\xi_{n\mathbf{k}}-\xi)
  \right\rangle}
  {2\left\langle \delta\left(\xi_{n\mathbf{k}}-\xi\right) \right\rangle}
\end{align}
is computed as an iso-energy average with the $\mathbf{k}$-dependent density of states as a weight.
The auxiliary energy-dependent gap function satisfies $\Delta_{n\mathbf{k}}^{(\mathbf{Q})}(\xi_{n\mathbf{k}}) = \Delta_{n\mathbf{k}}^{(\mathbf{Q})}$.

With the decoupling approximation for $\mathbf{Q}\neq0$, the current density in Eq.~(\ref{eq_cdensity}) becomes
\begin{align}
  \mathbf{j}^{(\mathbf{Q})}(\mathbf{r})=i\sum_{n\mathbf{k}}
  &\frac{1}{2E_{n\mathbf{k}}^{(\mathbf{Q})}}
  \left[\left\{
    \varphi_{n\mathbf{k}+\mathbf{Q}/2}^*(\mathbf{r})\boldsymbol{\nabla}
    \varphi_{n\mathbf{k}+\mathbf{Q}/2}(\mathbf{r})
   -\varphi_{n\mathbf{k}+\mathbf{Q}/2}^*(\mathbf{r})\boldsymbol{\nabla}
    \varphi_{n\mathbf{k}+\mathbf{Q}/2}(\mathbf{r})
  \right\}
  \left(E_{n\mathbf{k}}^{(\mathbf{Q})}+\bar{\xi}_{n\mathbf{k}}^{(\mathbf{Q})}\right)
  f\left(d_{n\mathbf{k}}^{(\mathbf{Q})}+E_{n\mathbf{k}}^{(\mathbf{Q})}\right)
  \right.
  \nonumber \\
  &\left. -\left\{
    \varphi_{n\mathbf{k}-\mathbf{Q}/2}^*(\mathbf{r})\boldsymbol{\nabla}
    \varphi_{n\mathbf{k}-\mathbf{Q}/2}(\mathbf{r})
   -\varphi_{n\mathbf{k}-\mathbf{Q}/2}^*(\mathbf{r})\boldsymbol{\nabla}
    \varphi_{n\mathbf{k}-\mathbf{Q}/2}(\mathbf{r})
  \right\}
  \left(E_{n\mathbf{k}}^{(\mathbf{Q})}-\bar{\xi}_{n\mathbf{k}}^{(\mathbf{Q})}\right)
  f\left(-d_{n\mathbf{k}}^{(\mathbf{Q})}-E_{n\mathbf{k}}^{(\mathbf{Q})}\right)
  \right]
\end{align}
To connect to the supercurrent density from the macroscopic theory represented in Eq.~(\ref{eq_j_macro}), we consider the spatially averaged current density
\begin{align}
  \bar{\mathbf{j}}^{(\mathbf{Q})} &\equiv
  \frac{1}{V}\int dr^3 \mathbf{j}^{(\mathbf{Q})}(\mathbf{r})
  \nonumber \\
  &=\frac{1}{V}\sum_{n\mathbf{k}}\frac{1}{E_{n\mathbf{k}}}
  \left[\mathbf{v}_{n\mathbf{k}+\mathbf{Q}/2}
    \left(E_{n\mathbf{k}}+\bar{\xi}_{n\mathbf{k}}\right)
   f\left(d_{n\mathbf{k}}+E_{n\mathbf{k}}\right)
   -\mathbf{v}_{n\mathbf{k}-\mathbf{Q}/2}\left(E_{n\mathbf{k}}-\bar{\xi}_{n\mathbf{k}}\right)
   f\left({-d}_{n\mathbf{k}}-E_{n\mathbf{k}}\right)
  \right].
\end{align}
Here, we used a partial integral
\begin{align}
  \frac{i}{2}\int dr^3
  \left\{
    \varphi_{n\mathbf{k}}^*(\mathbf{r})\boldsymbol{\nabla}
    \varphi_{n\mathbf{k}}(\mathbf{r})
   -\varphi_{n\mathbf{k}}^*(\mathbf{r})\boldsymbol{\nabla}
    \varphi_{n\mathbf{k}}(\mathbf{r})
  \right\}
  =\mathbf{\boldsymbol{\nabla}}_\mathbf{k}\xi_{n\mathbf{k}}\equiv \textbf{v}_{n\mathbf{k}}.
\end{align}
In the normal state ($\Delta_{n\mathbf{k}} = 0$), this quantity vanishes, 
\begin{align}
  \bar{\mathbf{j}}^{(\mathbf{Q})}\left(\Delta=0\right)
  =\frac{1}{V}\sum_{n\mathbf{k}}\mathbf{v}_{n\mathbf{k}+\mathbf{Q}/2}
  \left[
    1+\tanh\left(\frac{d_{n\mathbf{k}}+{\bar{\xi}}_{n\mathbf{k}}}{2T}\right)
  \right] = 0.
\end{align}
However, due to the finite spacing of the $\mathbf{k}$-point grid, a finite value remains.
Therefore, we subtract this residual from the total current as follows:
\begin{align}
  \bar{\mathbf{j}}_\textrm{sc}^{(\mathbf{Q})} &\equiv
  \bar{\mathbf{j}}^{(\mathbf{Q})} - 
  \bar{\mathbf{j}}^{(\mathbf{Q})}\left(\Delta=0\right)
  \\
  \bar{\mathbf{j}}_\textrm{sc}^{(\mathbf{Q})}
  &=\frac{1}{V}\sum_{n\mathbf{k}}\mathbf{v}_{n\mathbf{k}+\mathbf{Q}/2}
  \left[
    \frac{1}{2}
    \left\{
      \tanh\left(\frac{d_{n\mathbf{k}}+E_{n\mathbf{k}}}{2T}\right)
     +\tanh\left(\frac{d_{n\mathbf{k}}-E_{n\mathbf{k}}}{2T}\right)
    \right\}
  \right.
  \nonumber \\
  &\qquad\left.
    +\frac{{\bar{\xi}}_{n\mathbf{k}}}{2E_{n\mathbf{k}}}
    \left\{
      \tanh\left(\frac{d_{n\mathbf{k}}+E_{n\mathbf{k}}}{2T}\right)
     -\tanh\left(\frac{d_{n\mathbf{k}}-E_{n\mathbf{k}}}{2T}\right)
    \right\}
    -\tanh\left(\frac{d_{n\mathbf{k}}+{\bar{\xi}}_{n\mathbf{k}}}{2T}\right)
  \right].
\end{align}
Finally, we also use the auxiliary energy-grid scheme to stabilize the $\mathbf{k}$-integral as: 
\begin{align}
  \bar{\mathbf{j}}_\textrm{sc}^{(\mathbf{Q})}
  =\frac{1}{V}\int{d\xi}\sum_{n\mathbf{k}}
  \delta(\xi-{\bar{\xi}}_{n\mathbf{k}}^{(\mathbf{Q})})
  \mathbf{v}_{n\mathbf{k}+\mathbf{Q}/2}
  \left[
    \frac{1}{2}
    \left\{
      \tanh\left(\frac{d_{n\mathbf{k}}(\xi)+E_{n\mathbf{k}}(\xi)}{2T}\right)
     +\tanh\left(\frac{d_{n\mathbf{k}}(\xi)-E_{n\mathbf{k}}(\xi)}{2T}\right)
    \right\}
  \right.
  \nonumber \\
  \left.
    +\frac{\xi}{2E_{n\mathbf{k}}(\xi)}
    \left\{
      \tanh\left(\frac{d_{n\mathbf{k}}(\xi)+E_{n\mathbf{k}}(\xi)}{2T}\right)
     -\tanh\left(\frac{d_{n\mathbf{k}}(\xi)-E_{n\mathbf{k}}(\xi)}{2T}\right)
    \right\}
    -\tanh\left(\frac{d_{n\mathbf{k}}(\xi)+\xi}{2T}\right)
  \right].
\end{align}
\section{Result}
\begin{table}[!tb]
\caption{Cutoff energy used to expand $\varphi_{n\mathbf{k}}(\mathbf{r})$ with plane waves, $\mathbf{q}$-point grid for phonon and response function calculations, and the pseudopotential library used for each system.
PSL, GBRV, and SG15 indicate Pslibrary~\cite{DALCORSO2014337}, Garrity--Bennett--Rabe--Vanderbilt high-throughput pseudopotentials~\cite{GARRITY2014446}, and Schlipf--Gygi optimized norm-conserving pseudopotentials~\cite{SCHLIPF201536,Scherpelz2016}, respectively.
Experimental and theoretical lattice constants together with the structure type are also shown.
The lattice constants of elemental bulks are from Ref.~\cite{ashcroft1976solid}.
\label{tab_condition}}
\begin{center}
\begin{tabular}{ccccccccc}
  \hline
material & Al & In & Sn & Ta & Pb & Nb & H$_3$S & V$_3$Si \\
   \hline
 cutofffor$\varphi$~(Ry) &30&50&60&45&35&40&60& 35 \\
 $\mathbf{q}$ grid & $8^3$ & $7^3$ & $5^3$ & $8^3$ & $6^3$ & $8^3$ & $9^3$ & $5^3$ \\
 pseudopotential & PSL-PAW & PSL-US & GBRV-US & GBRV-US & SG15-NC & PSL-PAW & \begin{tabular}{c}H:PSL-US\\S:GBRV-US\end{tabular} & \begin{tabular}{c}V:GBRV-US\\Si:PSL-US\end{tabular} \\
   \hline
 structure & fcc & bct & $\beta$-Sn & bcc & fcc & bcc & Im$\bar{3}$m & A15 \\
\rowcolor{gray!10}
 $a_\textrm{exp}$(\AA) &4.05&5.82&4.59&3.31&4.95&3.3& 3.01~\cite{Einaga2016} & 4.719~\cite{STAUDENMANN1978461} \\
 $a_\textrm{calc}$(\AA) &4.04&5.95&4.7&3.31&5.05&3.31&2.99& 4.69 \\
\rowcolor{gray!10}
 $c_\textrm{exp}$(\AA) & NA &3.17&4.94& NA & NA & NA & NA & NA \\
 $c_\textrm{calc}$(\AA) & NA &3.22&4.99& NA & NA & NA & NA & NA \\
   \hline
\end{tabular}
\end{center}
\end{table}
We performed benchmark calculations on elemental bulk materials (Al, Nb, Sn, In, Pb, Ta), an A15 compound V$_3$Si, and a hydrogen compound H$_3$S under high pressure (200 GPa).

To calculate the normal-state atomic, electronic, and phononic structures as a preprocessing step for the construction of the SCDFT gap equation, we employed the first-principles program package {\sc Quantum ESPRESSO}~\cite{Giannozzi_2017}, which describes Kohn--Sham orbitals and atomic potentials using plane waves and pseudopotentials.
The cutoff plane-wave energies and the choice of projector-augmented wave (PAW)~\cite{PhysRevB.50.17953}, ultrasoft (US)~\cite{PhysRevB.41.7892}, and norm-conserving (NC)~\cite{PhysRevB.88.085117} pseudopotentials for each system, as summarized in Table~\ref{tab_condition}, were determined by verification with the Standard Solid State Pseudopotential dataset~\cite{Prandini2018}.
The Perdew--Burke--Ernzerhof (PBE) form of the generalized gradient approximation was used to construct the exchange correlation energy functional.
We computed phonons and electron-phonon interactions using density functional perturbation theory (DFPT)~\cite{RevModPhys.73.515}.
The Bloch wavenumber ($\mathbf{q}$) grid for phonons is proportional to the size of the Brillouin zone (see Table~\ref{tab_condition}).
The $\mathbf{k}$-point grids of electronic states for structure optimization, charge self-consistent calculations, and DFPT calculations are twice as fine as the $\mathbf{q}$ grid in each direction.
For the calculation of Pb, we included spin-orbit interaction (SOI) because this interaction substantially enhances electron-phonon interaction~\cite{PhysRevB.81.174527}.
We also provide the result without SOI to see the impact of that effect.

SCDFT calculations were performed using the open-source program package Superconducting-Toolkit (SCTK)~\cite{PhysRevB.101.134511}.
We treated electron-phonon correlation using the Eliashberg-combined SCDFT functional~\cite{PhysRevLett.125.057001}.
The plasmon-assisted kernel~\cite{PhysRevLett.111.057006} and mass renormalization~\cite{PhysRevB.102.214508} were computed using the random-phase approximation, while the magnetic exchange-correlation kernel in the spin-fluctuation-mediated interaction~\cite{PhysRevB.90.214504} was treated with the adiabatic local density approximation.
The Bloch wavenumber ($\mathbf{q}$) grid for the dynamical magnetic and electronic response functions was the same as that used for phonons, while the explicitly energy-dependent part (Lindhard's function) was computed on a grid four times denser, together with the optimized tetrahedron method~\cite{PhysRevB.89.094515}.
To compute the response function and the gap equation, we included $20\times N_\textrm{atom}$ empty bands per spin, where $N_\textrm{atom}$ is the number of atoms in the unit cell.

\begin{figure}[tbp!]
\includegraphics[width=1\textwidth]{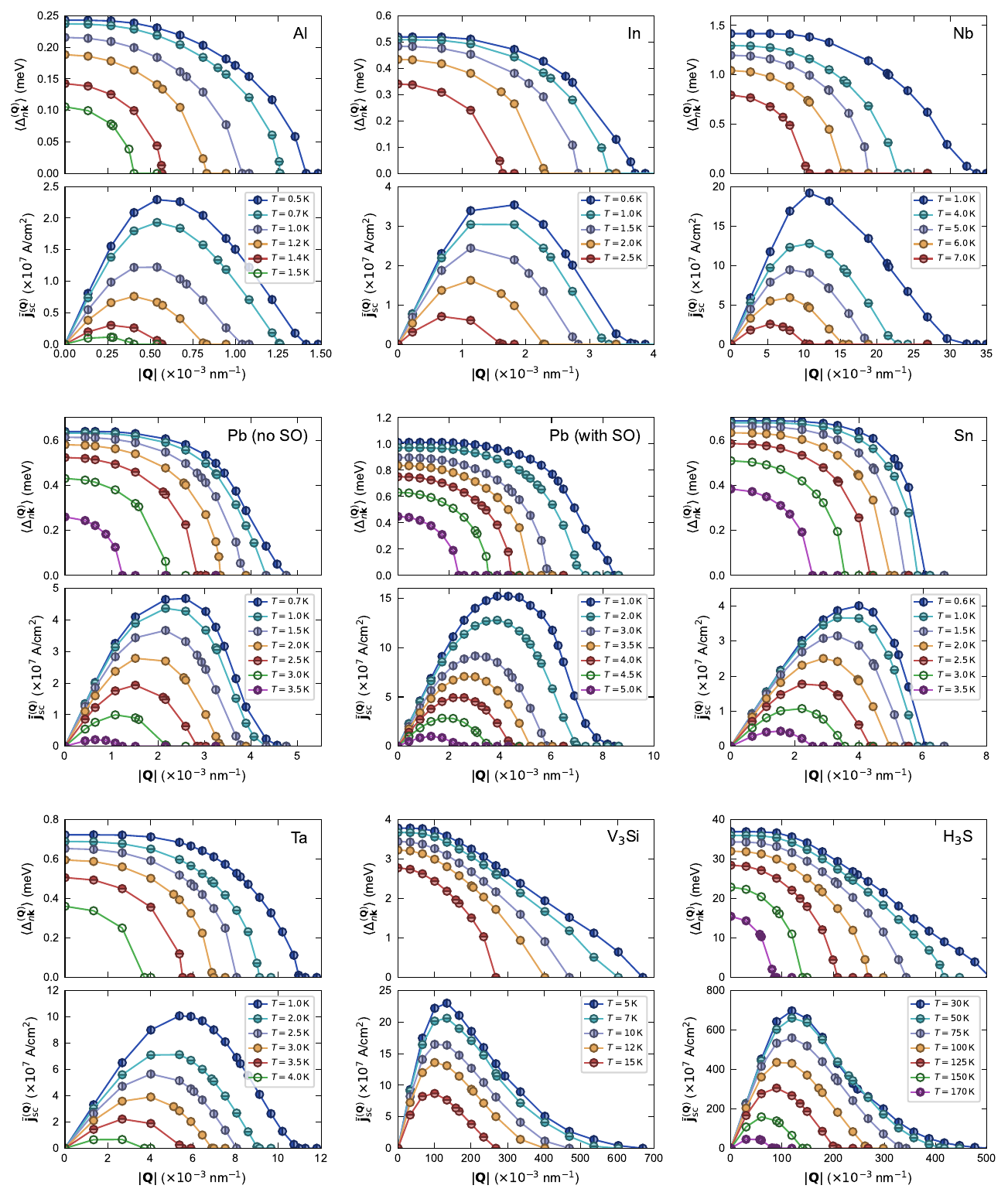}
\caption{\label{fig:deltaq_all} 
(Upper panel of each material) Fermi-surface-averaged superconducting gap function, $\langle\Delta_{n\mathbf{k}}^{(\mathbf{Q})}\rangle$, plotted as a function of the Cooper pair momentum.
(Lower panel of each material) Supercurrent plotted as a function of $Q$.
}
\end{figure}

\begin{figure}[tbp!]
\includegraphics[width=1\textwidth]{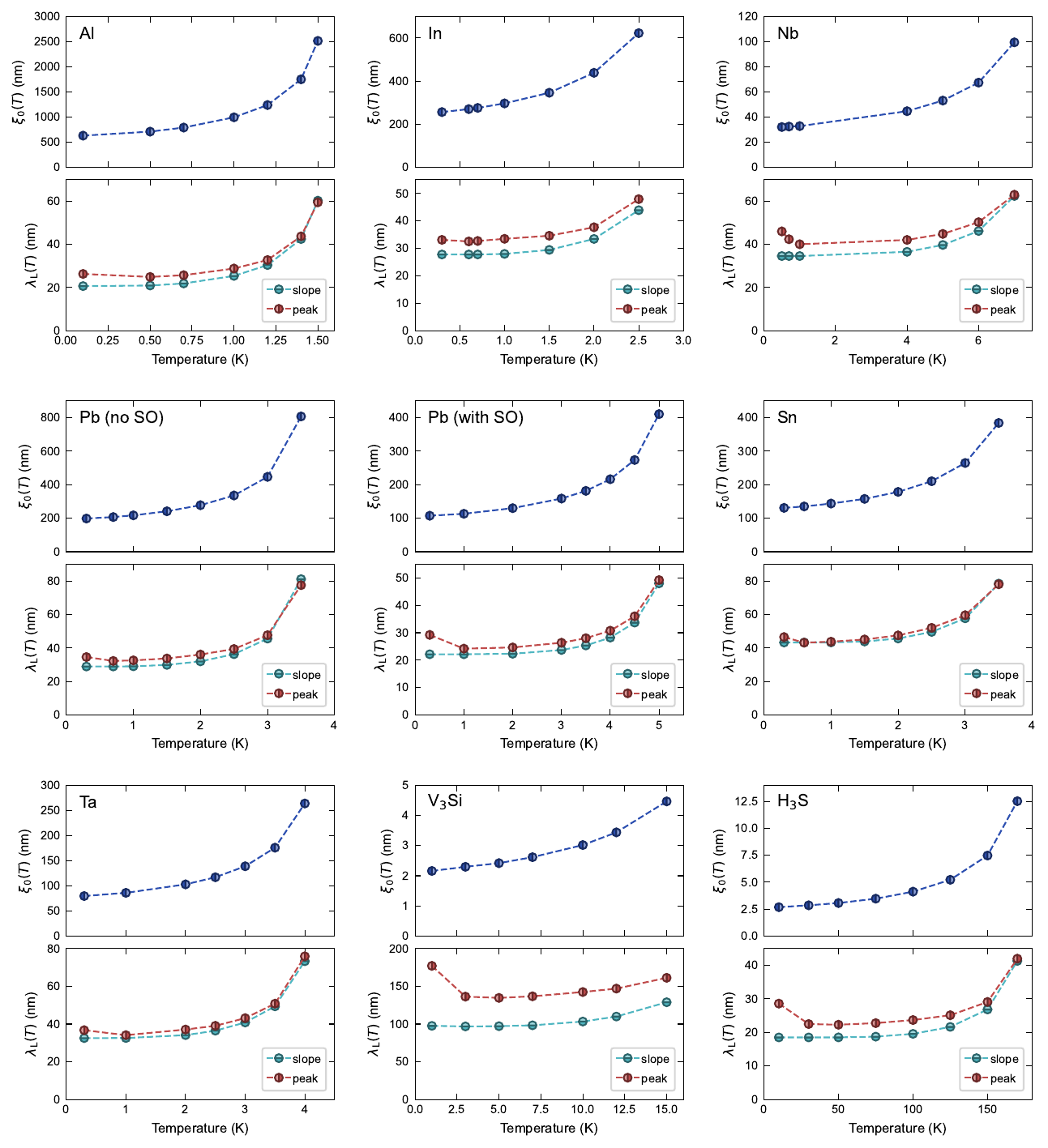}
\caption{\label{fig:xi_lam_all}
(Upper panel of each material) Temperature dependence of the calculated coherence length $\xi_0$. 
(Lower panel of each material) Temperature dependence of the penetration depth $\lambda_\mathrm{L}$, calculated from both the peak and the slope of the supercurrent.
}
\end{figure}

Table~\ref{table_calculation} summarizes the calculated and experimental quantities.
Inside that table, we also listed the penetration depth estimated by the normal-state quantities as:
\begin{align}
\lambda_\textrm{L,n} \equiv \sqrt{\frac{3(1+\lambda_\mathrm{ep})}{\mu_0 D(\varepsilon_\textrm{F}) \langle|\mathbf{v}_{n\textbf{k}}|^2\rangle}}.
\end{align}
The coherence length and penetration depth were computed from the momentum ($\mathbf{Q}$) dependence of the averaged gap, $\langle\Delta_{n\mathbf{k}}^{(\mathbf{Q})}\rangle$, and the supercurrent, $\bar{\mathbf{j}}_\mathrm{sc}^{(\mathbf{Q})}$, as shown in Fig.~\ref{fig:deltaq_all}.
Fig.~\ref{fig:xi_lam_all} shows the temperature dependence of the coherence length and penetration depth. We note that for the penetration depth, the slope-based value [Eq.~(20) of the main text] decreases monotonically, whereas the peak-based value [Eq.~(21)] lies consistently above it and, for some materials, turns upward at the lowest temperatures. Such deviations are expected as Eq.~(21) rests on the Ginzburg--Landau relation and is accurate only near $T_{\mathrm{c}}$, whereas the slope-based Eq.~(20) is a direct linear-response result. The difference between the two, including the low-temperature upturn, grows at lower temperature and reflects the approximation underlying Eq.~(21) rather than a physical feature.
\newpage
\begin{table}[hp!]
\caption{Experimental and calculated values for the transition temperature ($T_\mathrm{c}$), coherence length ($\xi_0$), penetration depth ($\lambda_\mathrm{L}$), and their ratio.
We computed $\lambda_\mathrm{L}$ using two methods: from the peak (depairing current $J_\mathrm{dp}$) of the supercurrent [Eq.~(17)] and from the slope of that current [Eq.~(16)].
Experimental critical current $J_\mathrm{c}$, calculated depairing current $J_\mathrm{dp}$, Pippard's estimation of the coherence length with and without electron-phonon mass renormalization, the London penetration depth estimated from the normal-state properties, gap function $\langle\Delta_{n\textbf{k}} \rangle$ and Fermi velocity $\langle|\mathbf{v}_{n\textbf{k}} |\rangle$ averaged on Fermi surfaces, and the Fr\"ohlich electron-phonon mass-enhancement factor $\lambda_\mathrm{ep}$ are also shown.
The data in Refs.~\onlinecite{HUEBENER1990605,orlando1991foundations,van1981principles} were obtained from Ref.~\onlinecite{poole2014superconductivity}.
\label{table_calculation}
}
\begin{center}
\begin{tabular}{ccccccccc}
  \hline
material & Al & In & Sn & Ta & Pb & Nb & H$_3$S & V$_3$Si \\
  \hline
\rowcolor{gray!10}
Exp. $T_\mathrm{c}$ (K) & 1.140~\cite{kittel2004introduction} & 3.4035~\cite{kittel2004introduction} & 3.722~\cite{kittel2004introduction} & 4.483~\cite{kittel2004introduction} & 7.193~\cite{kittel2004introduction} & 9.50~\cite{kittel2004introduction} & 184~\cite{Drozdov2015} & 17.1~\cite{kittel2004introduction} \\
Calc. $T_\mathrm{c}$ (K) &1.61&3.06&4.02&4.42& \begin{tabular}{c}5.33\\3.74(noSOI)\end{tabular} &8.09&181& 20.5 \\
\rowcolor{gray!10}
Exp. $\xi_0$ (nm) & \begin{tabular}{c}500--1,600~\cite{buckel1991superconductivity}\\1,600~\cite{HUEBENER1990605,orlando1991foundations}\end{tabular} & \begin{tabular}{c}260~\cite{PhysRev.123.442}\\275~\cite{HUEBENER1990605}\\360~\cite{orlando1991foundations}\\360--440~\cite{buckel1991superconductivity}\\440~\cite{van1981principles}\end{tabular} & \begin{tabular}{c}94~\cite{HUEBENER1990605}\\120--320~\cite{buckel1991superconductivity}\\230~\cite{orlando1991foundations,van1981principles}\end{tabular} & \begin{tabular}{c}92.5(4)~\cite{PhysRevLett.22.229}\\92.5~\cite{PhysRevB.7.136}\\93~\cite{buckel1991superconductivity}\end{tabular} & \begin{tabular}{c}51--83~\cite{buckel1991superconductivity}\\74~\cite{HUEBENER1990605}\\83~\cite{kittel2004introduction,van1981principles}\\90(5)~\cite{PhysRevB.72.024506}\\90~\cite{orlando1991foundations}\end{tabular} & \begin{tabular}{c}38~\cite{kittel2004introduction,van1981principles}\\39--40~\cite{buckel1991superconductivity}\\39.9(25)~\cite{10.1103/2nsw-n8gf}\\40~\cite{orlando1991foundations}\end{tabular} & 2~\cite{Mozaffari2019} & \begin{tabular}{c}3~\cite{orlando1991foundations}\\4~\cite{Zehetmayer_2014,buckel1991superconductivity}\end{tabular} \\
Calc. $\xi_0$ (nm) &636&254&132&80& \begin{tabular}{c}104\\198(noSOI)\end{tabular} &34&3& 2.2 \\
$\xi_\textrm{Pippard}$ (nm) &1011&517&325.4&161.9&223&79.6&5.453& 7.575 \\
$\xi_\textrm{Pippard,ep}$ (nm) &712&264&168&82.2&89.7&34&1.46& 1.74 \\
\rowcolor{gray!10}
Exp. $\lambda_\textrm{L}$ (nm) & \begin{tabular}{c}15.7~\cite{Lopez-Nunez_2025}\\16~\cite{kittel2004introduction,orlando1991foundations}\\50~\cite{buckel1991superconductivity,HUEBENER1990605}\end{tabular} & \begin{tabular}{c}21~\cite{van1981principles}\\24--64~\cite{buckel1991superconductivity}\\29.9~\cite{Raychaudhuri1985}\\47~\cite{HUEBENER1990605}\\65~\cite{orlando1991foundations}\end{tabular} & \begin{tabular}{c}25--50~\cite{buckel1991superconductivity}\\34~\cite{kittel2004introduction}\\36~\cite{van1981principles}\\50~\cite{orlando1991foundations}\\52~\cite{HUEBENER1990605}\end{tabular} & \begin{tabular}{c}33(1)~\cite{PhysRevLett.22.229}\\34.5~\cite{PhysRevB.7.136}\\35~\cite{buckel1991superconductivity}\end{tabular} & \begin{tabular}{c}32--39~\cite{buckel1991superconductivity}\\37~\cite{kittel2004introduction,van1981principles}\\40~\cite{orlando1991foundations}\\47~\cite{HUEBENER1990605}\end{tabular} & \begin{tabular}{c}27(3)~\cite{PhysRevB.72.024506}\\29.1(10)~\cite{10.1103/2nsw-n8gf}\\32--44~\cite{buckel1991superconductivity}\\39~\cite{kittel2004introduction,van1981principles}\\85~\cite{orlando1991foundations}\end{tabular} & \begin{tabular}{c}20~\cite{Minkov2022}\\37~\cite{Minkov2023}\end{tabular} & \begin{tabular}{c}60~\cite{orlando1991foundations}\\70~\cite{buckel1991superconductivity}\\90~\cite{Zehetmayer_2014}\end{tabular} \\
Calc. $\lambda_\textrm{L}$ from peak (nm) &26&33&43&34& \begin{tabular}{c}24\\33(noSOI)\end{tabular} &40&22& 136 \\
Calc. $\lambda_\textrm{L}$ from slope (nm) &21&28&43&33& \begin{tabular}{c}22\\29(noSOI)\end{tabular} &34&19& 97 \\
$\lambda_\textrm{L,n}$ (nm) &21&30&36&34&40&37&34&140 \\
\rowcolor{gray!10}
Exp. $\lambda_{\mathrm{L}}/\xi_0$ & 0.0098--0.1 & 0.05--0.25 & 0.11--0.55 & 0.35--0.38 & 0.36--0.92 & 0.68--2.24 & 10--19 & 15--30 \\
Calc. $\lambda_{\mathrm{L}}/\xi_0$ & 0.033--0.041 & 0.11--0.13 &0.33& 0.41--0.43 & 0.21--0.23 & 1.0--1.2 & 6.3--7.3 & 44--62 \\
\rowcolor{gray!10}
Exp.~$J_\mathrm{c}$ ($\times10^7$ A/cm$^2$) & 1.1--2.5~\cite{PhysRevB.26.3648} & 0.88--1.1~\cite{PhysRevB.1.4295} & 0.5--0.8~\cite{PhysRevB.1.4295} & 0.01145~\cite{HU201885} & NA & 0.75~\cite{PhysRevB.89.184511} & NA & 0.01~\cite{Zehetmayer_2014} \\
Calc.~$J_\mathrm{dp}$ ($\times10^7$ A/cm$^2$) &2.3&3.6&4&10& \begin{tabular}{c}15\\4.7(noSOI)\end{tabular} &19&697& 24 \\
$\langle\Delta_{n\textbf{k}}\rangle$ (meV) &0.295&0.581&0.732&0.766&1.081&1.466&38.77& 3.955 \\
$\langle|\mathbf{v}_{n\textbf{k}}|\rangle$ ($\times10^6$ m/s) &1.423&1.433&1.137&0.592&1.152&0.557&1.009& 0.143 \\
$\lambda_\mathrm{ep}$ &0.42&0.96&0.94&0.97&1.49&1.34&2.73& 3.36 \\
\hline
\end{tabular}
\end{center}
\end{table}

\newpage
\bibliography{ref}